\documentclass[10pt,final]{IEEEtran}
\usepackage{cite}

\ifCLASSINFOpdf
  \usepackage[pdftex]{graphicx}
\else
  \usepackage[dvips]{graphicx}
\fi
\usepackage{amssymb}
\usepackage[cmex10]{amsmath,mathtools}
\usepackage{bm}
\usepackage{mathabx}
\usepackage{nccmath}
\usepackage{hyperref}                 
\usepackage{algorithm}
\usepackage{algpseudocode}

\usepackage{array}

\usepackage{mdwmath}
\usepackage{mdwtab}

\usepackage{epstopdf}
\usepackage[tight,footnotesize]{subfigure}

\usepackage{color,soul}

\usepackage{xcolor}

\usepackage{siunitx}
\usepackage{setspace}
\newcommand{\vect}[1]{\mathbf{#1}}
\usepackage{hhline}
\newtheorem{theorem}{Theorem}[section]
\newtheorem{definition}[theorem]{Definition}

\newenvironment{defn*}{\begin{definition}}{\end{definition}}

\def\approxprop{%
  \def\p{%
    \setbox0=\vbox{\hbox{$\propto$}}%
    \ht0=0.6ex \box0 }%
  \def\s{%
    \vbox{\hbox{$\sim$}}%
  }%
  \mathrel{\raisebox{0.7ex}{%
      \mbox{$\underset{\s}{\p}$}%
    }}%
}

\begin{document}
%
\title{Coherent Track Before Detect: Detection via simultaneous trajectory estimation and long time integration}
%
%
%

\author{Kimin~Kim,~\IEEEmembership{Student Member,~IEEE},
        Murat~\"{U}ney,~\IEEEmembership{Member,~IEEE},
        and~Bernard~Mulgrew,~\IEEEmembership{Fellow,~IEEE}
\thanks{This work was supported by the Engineering and Physical Sciences Research Council (EPSRC) grant EP/K014277/1, and, the MOD University Defence Research Collaboration (UDRC) in Signal processing. 

K. Kim, M. \"{U}ney, and B. Mulgrew are with Institute for Digital Communications, School of Engineering, the University of Edinburgh, Edinburgh, EH9 3JL, U.K. (e-mail:\{K.Kim, M.Uney, B.Mulgrew\}@ed.ac.uk).}}
\maketitle

\begin{abstract}
In this work, we consider the detection of manoeuvring small objects with radars. Such objects induce low signal to noise ratio (SNR) reflections in the received signal. We consider both co-located and separated transmitter/receiver pairs, i.e., mono-static and bi-static configurations, respectively, as well as multi-static settings involving both types. We propose coherent track before detect: A detection approach which is capable of coherently integrating these reflections within a coherent processing interval (CPI) in all these configurations and continuing integration for an arbitrarily long time across consecutive CPIs. {We estimate the complex value of the reflection coefficients for integration while simultaneously estimating the object trajectory. Compounded with these computations is the estimation of the unknown time reference shift of the separated transmitters necessary for coherent processing.} Detection is made by using the resulting integration value in a Neyman-Pearson test against a constant false alarm rate threshold. We demonstrate the efficacy of our approach in a simulation example with a very low SNR object which cannot be detected with conventional techniques.
\end{abstract}


%
\IEEEpeerreviewmaketitle

\section{Introduction}
\IEEEPARstart{T}{he} detection of manoeuvring and small objects with radars is a challenging task~\cite{RichardsMelvinScheerEtAl2014} and is a highly desirable capability in surveillance applications~\cite{Haykin2006}. Radars emit modulated pulses towards a surveillance region and collect reflected versions of the transmitted waveforms from objects in this area. Small objects induce low signal-to-noise ratio (SNR) signals at the radar receiver. The decision on object presence is made by testing the hypothesis that the received signal contains reflections against the noise only signal hypothesis after the front-end input is filtered with a system response matching the probing waveform, which is known as the matched filter~(MF)~\cite{Richards2005b}.   

In order to detect low SNR objects, many such pulse returns (i.e., multiple measurements) need to be considered as each reflection is at a level similar to the noise background. The sufficient statistics of multiple pulse returns are found by summing the associated reflection coefficients across them, which is referred to as pulse integration~\cite[Chp.8]{Richards2005b}. This process is applied on the sampled outputs of the MF stage. These samples correspond to, in effect, measurements corresponding to resolution bins in an equally divided range space. In conventional processing, beam-forming and Doppler processing with these samples are used to further segment the bearing and Doppler space into resolution bins and find the corresponding measurements. Conventional methods for integration over time such as coherent and non-coherent integration integrate pulse returns in the same range-bearing and Doppler bins across time. When objects manoeuvre, however, these reflections follow a trajectory across these bins, and, these methods fail to collect evidence on object existence for a long time due to not taking into account this trajectory. On the other hand, longer integration time provides higher probability of detection for a given false alarm rate, in principle. 

One possible solution to providing long time integration for manoeuvring objects is to design filters with long impulse responses that match multiple pulse returns along a selection of possible trajectories (see, e.g.,~\cite{Chen2014}\&\cite{KongLiCuiEtAl2015}). The number of filters required in this approach easily becomes impractically excessive with increasing integration time. An alternative approach is to employ a dynamic programming perspective and use a regular probing pulse MF to integrate its outputs along a trajectory estimated simultaneously which corresponds, in a sense, to on-line adaptive synthesis of long time MFs. 

Trajectory estimation using the outputs of a pulse MF is often referred to as track-before-detect (see, for example, \cite{Boers2004,Grossi2013a}). The sample that corresponds to the true object kinematic state (i.e., location and velocity) is a complex value that is a sum of the reflection coefficient and background noise~\cite{VanTrees1992b}
. Most track-before-detect algorithms, on the other hand, use the modulus of the MF within models which describe the statistics of the modulus of the MF output. These models are averaged and hence cannot fully exploit the information captured by the measurements. For example, it is well known that the detection performance of these methods can be improved by also taking into account the phase of the data samples~\cite{Davey2012}, in addition to the modulus. 

The best achievable detection performance is obtained by coherent processing~\cite{Richards2005b}, in which one needs to estimate the complex reflection coefficient from the complex values of the MF outputs, the latter of which are processed by the aforementioned algorithms. This corresponds to using a non-averaged model in which the reflection coefficient is a random variable that remains the same during what is known as a coherent processing interval (CPI), and, is generated randomly for consecutive CPIs~\cite{VanTrees1992b}. This is challenging partly because estimation of this quantity with a reasonable accuracy requires more samples than one can collect at the pulse-width sampling rate in a coherent processing interval (CPI)~\cite{Uney2015}. For example, in~\cite{Rabaste2012}, coherent processing and integration within a CPI is performed with a very high sampling rate that yields a large number of samples in the pulse interval.

In~\cite{KimUneyMulgrew2016}, we demonstrated that this can be remedied using a phased array receiver structure. In particular, we introduced coherent track before detect: A simultaneous trajectory estimation and long time integration algorithm in which the integrated value is then tested against a constant false alarm rate (CFAR) threshold for declaring the existence or otherwise of an object in a Neyman-Pearson sense. In~\cite{KimUeneyMulgrew2017}, we extend this approach for separated transmitter/receiver pairs, i.e., bi-static channels, with an unknown time reference shift. We recover the synchronisation term by diverting simultaneous beams towards the tested point of detection and the remote transmitter thereby relaxing the commonly used assumption that the remote transmitters and the local receiver are synchronised (see, e.g., \cite{Godrich2010, Niu2012}).

\begin{figure}[t]
	\centering
	\includegraphics[width=3in]{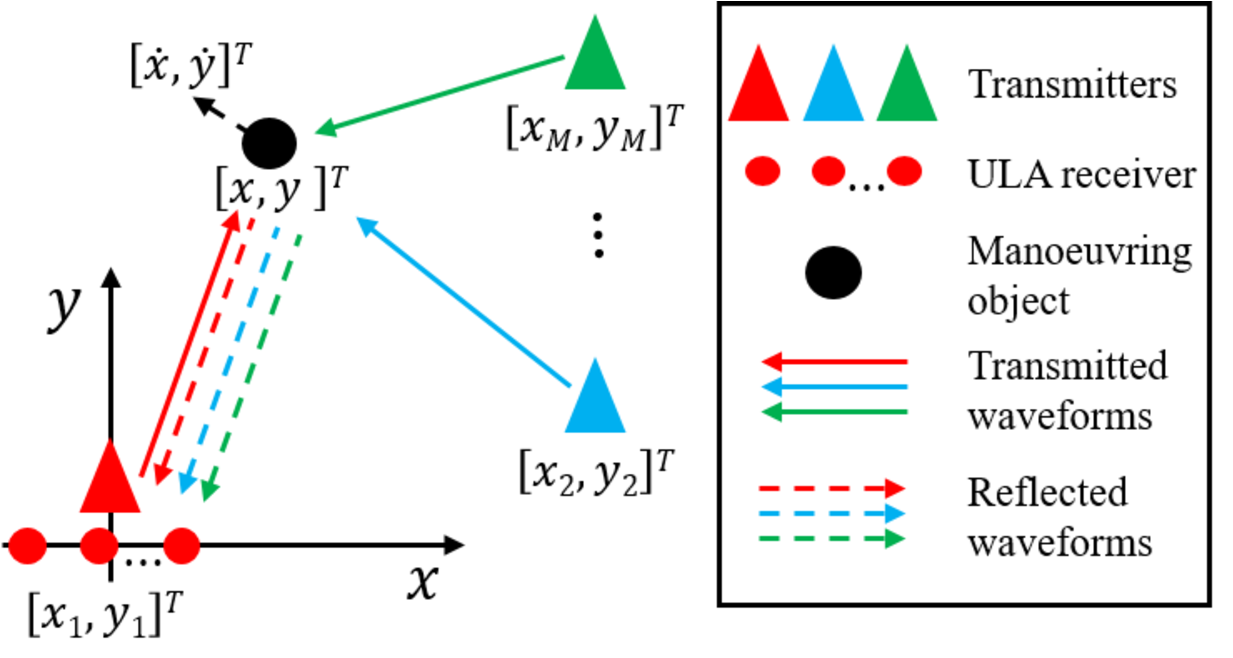}
	\centering
	\vspace{-0.5em}
	\caption{Problem scenario: $M$ transmitters and a ULA receiver to detect a small object located at $[x,y]^T$ with velocity $[\dot{x},\dot{y}]^T$.}
	\label{fig:multiradar}
\end{figure}

In this work, we provide a complete exposition of our coherent track before detect, equivalently, long time integration and trajectory estimation approach in mono-static and bi-static configurations as well as the multi-static case. 
In particular, we consider the system structure in \figurename~\ref{fig:multiradar} where there are multiple transmitters using mutually orthogonal waveforms. The receiver is a ULA and has the full knowledge of the transmission characteristics except the time reference shift of the separately located transmitters. The front-end signals at the receive elements are the superposition of noise, signals from direct channels, and, reflections from objects.

We consider a long time likelihood ratio test conditioned on a trajectory in a kinematic state space, reflection coefficients, and, synchronisation terms as unknown parameters. In order to estimate the kinematic quantities, we use a Markov state-space model in which the object state consists of location and velocity variables. The measurement model of this state space model involves the radar ambiguity function parametrised on the aforementioned reflection coefficients. These coefficients are estimated by using an expectation-maximisation algorithm~\cite{Moon1996a} realising a maximum likelihood (ML) approach within Bayesian filtering recursions for state trajectory estimate. We show that this is an empirical Bayesian method~\cite{CarlinLouis2010} for realising the update stage of the filter. When these ML estimates are reasonably accurate, the empirical Bayes update is an accurate approximation to the otherwise intractable filtering update equations. For synchronisation, we employ a digital beam-forming technique to simultaneously divert beams towards both the test points of detection and the locations of the separately located transmitters in order to find the respective time reference shifts in the bi-static channels.



The resulting algorithm enables us to collect the entire evidence of object existence at the receiver by i)~performing coherent integration in both mono-static and bi-static channels within a CPI, ii)~non-coherently integrating across different (non-coherent) channels, e.g., local mono-static and remote bi-static channels, and, iii)~continuing integration for an arbitrarily long interval that contains many CPIs. As a result, this approach enables us to detect manoeuvring and low SNR objects which cannot be detected using other techniques.

This article is organised as follows: Section~\ref{sec:problem_statement} gives details of the problem scenario and introduces the mathematical statement of the problem. In Section~\ref{sec:Simultaneous_tracking_and_time_integration}, we discuss trajectory estimation with the array measurements and detail the aforementioned empirical Bayes approach. In Section~\ref{sec:MLEstimation}, we first introduce an expectation-maximisation (EM) algorithm for the ML estimation of the complex reflection coefficients. Then, we detail the ML estimation of the synchronisation term. We combine these estimators and specify the proposed detection scheme in Section~\ref{sec:Long_time_integration}. The proposed detection algorithm is demonstrated in Section~\ref{sec:Example} in comparison with a clairvoyant detector and a conventional scheme in a scenario with a manoeuvring and very low SNR object. Finally, we conclude in Section~\ref{sec:Conclusion}.

\section{Problem statement}
\label{sec:problem_statement}

\begin{figure}[tb]
	\centering
	\includegraphics[width=3in]{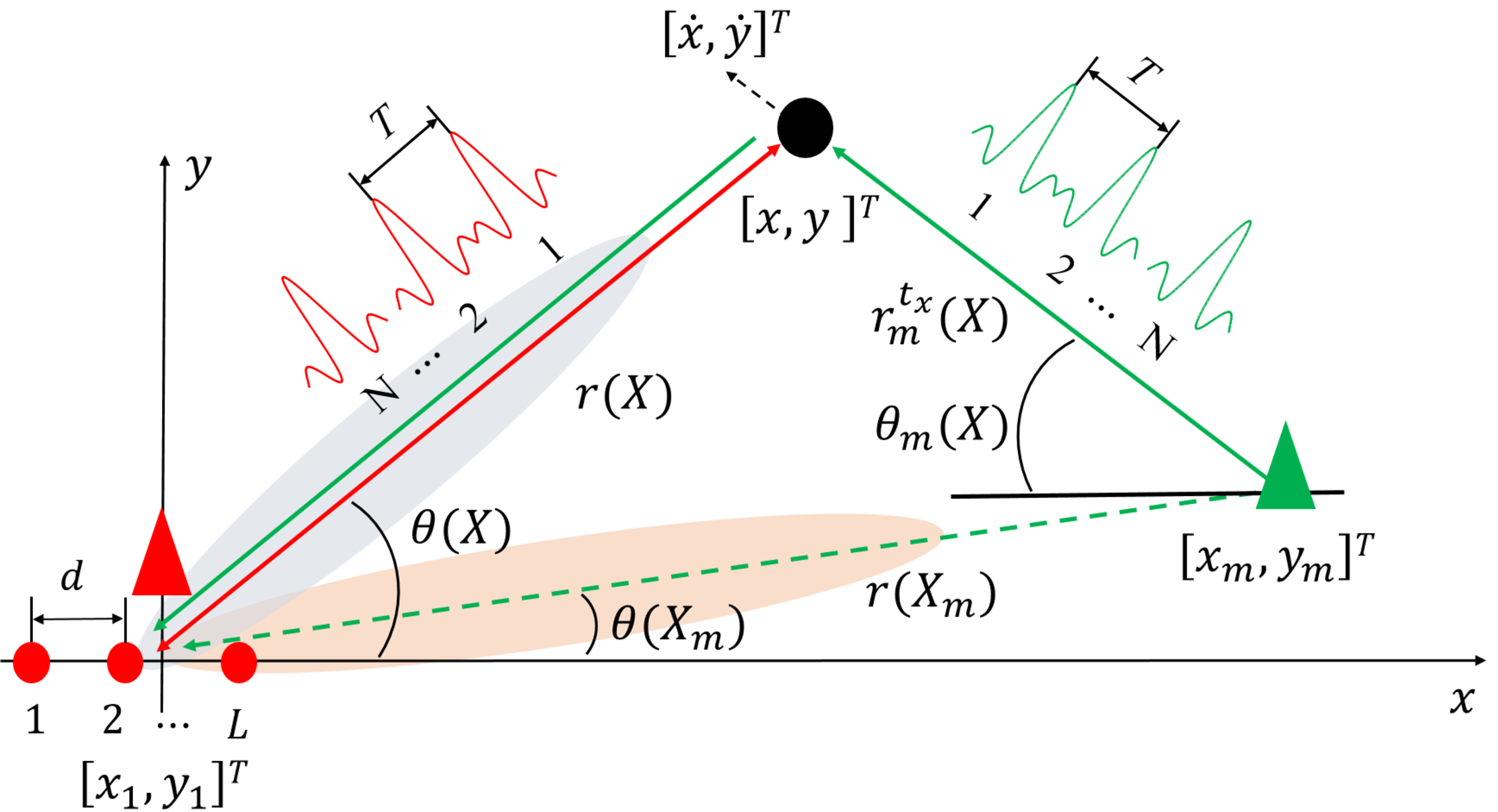}
	\centering
	\vspace{-0.5em}
	\caption{Geometry of the problem: A ULA receiver co-located with a transmitter and another transmitter placed in a separate location on the 2D Cartesian plane. Both polar and Cartesian coordinate variables are depicted. Each transmitter emits $N$ pulses in a CPI. The waveforms used are orthogonal.}
	\vspace{-1.5em}
	\label{fig:2txscenario}
\end{figure}

Let us consider the problem scenario in \figurename~\ref{fig:multiradar} with a ULA receiver (depicted by red dots), and, $M$ transmitters (depicted by triangles) one of which is co-located with the receiver forming a mono-static pair. The other transmitters are located elsewhere and form bi-static pairs with the receiver.

The receiver is comprised of $L$ elements spaced with a distance of $d$ which will be specified later in this section. Each element collects reflected versions of the transmitted waveforms emitted by both the co-located and the separately located transmitters thereby forming mono-static and bi-static pairs, respectively.
\subsection{Spatio-temporal signal model}
\label{sec:signal_model}
A detailed model for the signals induced at the receiver array by reflections from an object is as follows: We consider an interval of time in which each transmitter emits $N$ consecutive waveforms $\tilde u_m(t)$ separated by a time length of $T$ after modulating with a common carrier that has an angular frequency of $\omega_{c} = 2\pi f_c$. The $m$th transmitted signal is therefore given by 
\begin{equation}
u_{m}(t) = \mathrm{Re}{\left\{ \sum_{n=0}^{N-1} \tilde{u}_{m}(t-nT)\mathrm{e}^{j\omega_{c}t}\right\}},
\label{eqn:tranmitted pulse}
\end{equation}
where $\mathrm{Re}\{\cdot\}$ denotes the real part of its input complex argument and $T$ is known as the pulse repetition interval (PRI).

We assume that $\{\tilde{u}_{m}\}_{m=1}^M$ is an orthogonal set of waveforms of pulse duration $T_p$ and bandwidth $B$, i.e.,
\begin{eqnarray}
 <\tilde{u}_{m}(t), \tilde{u}_{m'}(t) > &\triangleq &\int_0^{T_p} \tilde{u}_{m}(t)\tilde{u}^*_{m'}(t)\mathrm{d}t \nonumber \\
 &=& \delta_{m,m'}
 \label{eqn:orthogonality}
\end{eqnarray}
for $m,m' \in \{1,\ldots,M\}$, where $\delta_{m,m'}$ is Kronecker's delta function. 

Use of such orthogonal transmit waveforms underlies the vision of multiple-input multiple-output (MIMO) radars~\cite{Haimovich2008,Li2009} a particular configuration of which is, hence, the system considered here. Design of orthogonal sets for MIMO sensing was investigated with various objectives such as maximisation of diversity~\cite{DeMaio2007} and waveform identifiability~\cite{YangBlum2007}. In this work, we consider a narrowband regime in which frequency division multiplexing can be used to achieve orthogonality in practice.

In order to specify the received signal at the array elements, let us consider the geometry of the problem which is illustrated in \figurename~\ref{fig:2txscenario} for ${M=2}$ transmitters. The receiver array measures the superposition of signals from different channels which are depicted by coloured lines. In particular, there are i)~a local (mono-static) channel (red line), ii)~a remote (bi-static) channel (green line), and, iii)~a direct channel from the remote transmitter (green dashed line). The first two are reflection channels propagating the reflected waveforms from the object (black circle) towards the receiver array. These channels can be fully separated given the array data by exploiting the orthogonality of the waveforms over time and the capability of spatial filtering thereby diverting multiple beams towards arbitrary arrival angles, simultaneously. These points will become clear in the sequel. 

Let us model the signals in the reflection channels. We assume that the reflectivity of the object remains coherent (i.e., unchanged) during the collection of reflections from the $N$ pulses in~\eqref{eqn:tranmitted pulse}. Such a time interval is known as a coherent processing interval (CPI). Modelling of the direct channel signals is introduced later in Section~\ref{sec:bistaticsynch}.

The kinematic state of the reflector (depicted by a black dot) in the 2D Cartesian plane is given by $X = [x,y,\dot{x},\dot{y}]^{T}$, where $[x,y]^{T}$ is the location, $[\dot{x},\dot{y}]^{T}$ is the velocity, and $T$ denotes vector transpose. The distance of $X$ to the receiver is related to pulse time of flights. The overall distance a pulse emitted by the $m$th transmitter at $[x_{m},y_{m}]^{T}$ and reaches the receiver at $[x_{1},y_{1}]^{T}$ after getting reflected at $[x,y]^T$ is given~by
\begin{eqnarray}
R_m(X) &=& R^{tx}_m(X) + R(X) \label{eqn:total_range} \\
R^{tx}_{m}(X) &\triangleq& \sqrt{(x - x_{m})^{2} + (y - y_{m})^{2}}. \nonumber \\
R(X) &\triangleq & \sqrt{(x - x_{1})^{2} + (y - y_{1})^{2}} \nonumber
\end{eqnarray}
where $R_{m}$ and $R$ denote the distance from the object to the $m$th transmitter and to the receiver, respectively.

The corresponding time of flights are found as
\begin{eqnarray}
\tau_{m}(X) = \tau^{tx}_m(X) + \tau(X)  && \label{eqn:tof} \\
\tau^{tx}_m(X) \triangleq \frac{R_{m}^{tx}(X)}{c}, &&\,\,\,\,\,\, \tau(X) \triangleq \frac{R(X)}{c}, \nonumber
\end{eqnarray}
where $c\approx \num{3.0e+8} m/s$ is the speed of light.

The velocity of the object induces an angular frequency shift on reflections which is known as the Doppler shift. This quantity is given by
\begin{eqnarray}
{\Omega_m}(X) =\frac{2\pi T}{\lambda_c}\Big(&\dot{x}&\times\left(\cos\theta(X)+\cos\theta_m(X)\right)\nonumber\\ &+&\dot{y}\times\left(\sin\theta(X)+\sin\theta_m(X)\right)\Big),
\label{eqn:doppler}
\end{eqnarray}
where $\theta$ and $\theta_m$ are the angle of arrival (AoA) of the reflections to the receiver and the bearing angle of the object with respect to the $m$th transmitter, respectively. These quantities are given by
\begin{eqnarray}
\theta(X) &= &\arctan(y_{1}-y)/(x_{1}-x) \nonumber \\
\theta_{m}(X) &= &\arctan(y_{m}-y)/(x_{m}-x).
\label{eqn:targetbearing}
\end{eqnarray}

For narrowband reflections, the signals induced at the array elements are characterised by a spatial steering vector as a function of $\theta$ which is given by \cite[Chp.2]{VanTrees2002m}  
\begin{equation}
\mathbf{s}_{s}(\theta) = \Big[1,\mathrm{e}^{-j\omega_c\frac{d}{c}\sin\theta}, \dots,
\mathrm{e}^{-j\omega_c(L-1)\frac{d}{c}\sin\theta} \Big]^T,
\notag
\end{equation}
where $d$ is the separation between the array elements selected as half of the carrier wavelength, i.e., $d=\lambda_c/2$. Substituting this quantity together with $c=\lambda_c \times f_c$ in the equation above leads to
\begin{equation}
\mathbf{s}_{s}(\theta) = \Big[1,\mathrm{e}^{-j\pi\sin\theta}, \dots,
\mathrm{e}^{-j(L-1)\pi\sin\theta} \Big]^T.
\label{eqn:ULA_model}
\end{equation}

The superposition of the reflections after demodulation at the receiver is given using~\eqref{eqn:ULA_model}~and~\eqref{eqn:doppler} by  
\begin{eqnarray}
\vect{z}(t) &=&\mathbf{s}_{s}(\theta) \sum_{m=0}^{M-1}\sum_{n=0}^{N-1}\alpha_{m}\mathrm{e}^{jn\Omega_m}\mathrm{e}^{-j\omega_{c}(\tau_{m}+\Delta t_{m})} \nonumber \\
&&\times\tilde{u}_{m}(t-\tau_{m}-\Delta t_{m}-nT),
\label{eqn:received_signal_in time}
\end{eqnarray}
where $\alpha_{m}$ is a complex coefficient modelling the reflectivity in the $m$th channel, and, $\tau_{m}$ is the time of flight of a pulse given in~\eqref{eqn:tof}. Here, $\Delta t_{m}$ is an unknown time shift modelling the time reference difference between the $m$th transmitter and the receiver (i.e., a synchronisation term).

The reflections in the received signal are optimally searched by matched filtering~\cite{VanTrees1992b}, i.e., by convolving the input with inverted versions of the probing waveforms. In our scenario, this corresponds to a bank of $M$ filters, (see, e.g. \cite[Chp.3]{Li2009}). Owing to the orthogonality (asserted by \eqref{eqn:orthogonality}), the $M$ channels in \eqref{eqn:received_signal_in time} will have been separated at the filter outputs\footnote{Perfect orthogonality of waveforms might not be achievable in practice, nevertheless, design of waveforms with a fairly small mutual cross-correlation has been a productive research area which is also discussed, for example, in \cite[Chp.2]{Li2009}.}. The output of the $m$th filter is given by
\begin{eqnarray}
 {\vect z}_m(t) &\triangleq& \vect{z}(t)*\tilde u_m(-t) \notag \\
 &=&\alpha_{m} \mathbf{s}_{s}(\theta) \sum_{n=0}^{N-1}\mathrm{e}^{jn\Omega_m}\mathrm{e}^{-j\omega_{c}(\tau_{m}+\Delta t_{m})}\notag \\
 &&\times \Lambda_{m}(t-\tau_{m}-\Delta t_{m}-nT).
\label{eqn:received_signal_matched_filter}
\end{eqnarray}
where $*$ denotes convolution and $\Lambda_m(\cdot)$ is the auto-correlation of the $m$th waveform given by  
\begin{eqnarray}
\Lambda_{m}(t) &=& \tilde u_m(t) * \tilde u_m(-t) \notag \\
&=& \int_0^{T_p}\tilde{u}_{m}(t')\tilde{u}^{*}_{m}(t'-t) \mathrm{d}t'.
\label{eqn:orthogonal_wave_auto} 
\end{eqnarray}

\begin{figure}[t]
	\centering
	\includegraphics[width=3in]{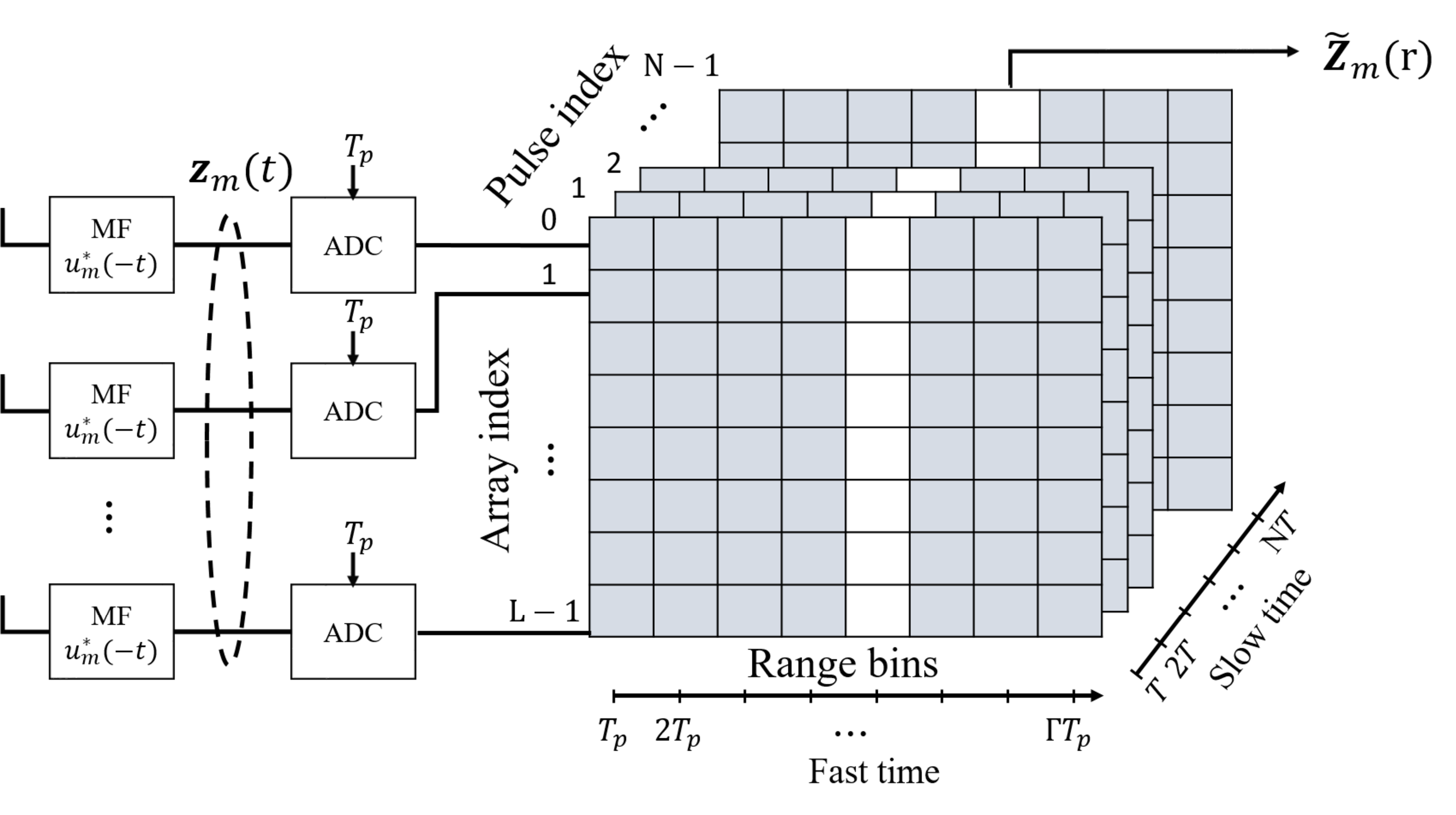}	
	\caption{Data acquisition in the $m$th channel: Sampled version of the received signal within a CPI as a radar data cube. The output of the matched filter is sampled and arranged in array index, fast time and slow time axis.}
	\vspace{-1.2em}
	\label{fig:Data_acquisition}
\end{figure}

This output is sampled with a period that equals to the pulse duration $T_p$. Let us assume that $T$ is an integer multiple of $T_p$, i.e., $T= \Gamma \times T_p $ where $\Gamma \in \mathbb{Z}^+$. $\Gamma \times N$ samples of this discrete time vector sequence is given by
\begin{eqnarray}
 \vect{z}_m[\gamma] &= &\vect{z}_m(\gamma T_p), \,\,\,\,\,\,\,\,\,\,\, \gamma = 1,\ldots,\Gamma \times N,
 \label{eqn:mthsampled} \\
&=& \alpha_m  s(\Delta t_m) \mathbf{s}_{s}(\theta)\mathbf{s}_{t}(\tau_{m},\Omega_m)^T  \nonumber \\
&& \times \left[ \begin{array}{c} \Lambda_{m}(\gamma T_p-\tau_{m}-\Delta t_{m}) \nonumber \\
\Lambda_{m}(\gamma T_p-\tau_{m}-\Delta t_{m} - T) \\
\vdots \\
\Lambda_{m}(\gamma T_p-\tau_{m}-\Delta t_{m} - (N-1)T)
\end{array} \right],
\end{eqnarray}
where 
\begin{eqnarray}
s(\Delta t_m ) &\triangleq &\mathrm{e}^{-j\omega_c \Delta t},
 \label{eqn:s_deltat} \\
  \mathbf{s}_{t}\left(\tau', \Omega' \right) &\triangleq& \mathrm{e}^{-j\omega_c \tau'} \times\Big[1,\mathrm{e}^{j\Omega'}, \dots,\mathrm{e}^{j (N-1) \Omega'}\Big]^T.
\label{eqn:temporal_model}
\end{eqnarray}
The term $ \mathbf{s}_{t}$ will be referred to as the temporal steering vector.

Next, this vector sequence is arranged as a cube by folding the two dimensional data array in \eqref{eqn:mthsampled} in lengths of $\Gamma$ samples. The $n$th layer of the resulting cube corresponds to the samples collected between the $n$th and the following pulse, i.e.,

\begin{equation}
 {\vect{C}}_m[n] \triangleq \Big[ \vect{z}_m[n\Gamma],\vect{z}_m[n\Gamma+1],\ldots,\vect{z}_m[(n+1)\Gamma-1]\Big] \notag
\end{equation}

This processing chain is illustrated in \figurename~\ref{fig:Data_acquisition} together with the cube ${\vect{C}}_m[n]$ which is also known as the radar data cube~\cite{Richards2005b}. The axes of this cube are array index, slow time and fast time. In the fast time axis, we have $\Gamma$ samples of the filter output, each of which is associated with a time delay of the reflected signal. These time delays correspond to time of flights which can easily be converted to range values using~\eqref{eqn:tof}. As a result, $N$ array measurements from range bin $r$ is a slice along the slow time axis given by
\begin{eqnarray}
 \vect{\tilde Z}_m(r) &\triangleq &\Big[ \vect{z}_m[ r ],\vect{z}_m[ \Gamma + r ],\cdots,\vect{z}_m[(N-1)\Gamma+ r ] \Big]  \nonumber \\
 &=& \alpha_m s(\Delta t_{m})\mathbf{s}_{s}(\theta) \mathbf{s}_{t}(\tau_{m},\Omega_m)^T \nonumber\\
 &&\times\Lambda_m( r T_p -\tau_{m} - \Delta t_{m}). \nonumber \\[-4pt]
 \label{eqn:measurement_vector}
\end{eqnarray}

For convenience regarding the notation in the rest of this article, we stack columns of $\vect{\tilde Z}_m(r)$ and form a $LN \times 1$ data vector. Before specifying this vector, let us combine the signal model in a single entity as a function of the reflector kinematic state $X$ which induces the signals and the range bin $r$ which is the measurement index:
\begin{eqnarray}
\mathbf{s}_{m}(r,X) &\triangleq& s(\Delta t_m)\mathbf{s}_{s}(\theta(X))\otimes \mathbf{s}_{t}( \tau_m(X),\Omega_m(X)) \nonumber\\
&&\times \Lambda_m( r T_p -\tau_{m}(X)-\Delta t_{m}) 
\label{eqn:column_data_model}
\end{eqnarray}
where $\otimes$ denotes the Kronecker product operator. Here, $X = [x,y,\dot{x},\dot{y}]$ is related to the data vector through the associated time of flight $\tau_m$ and $(R,\theta,\Omega_m)$ found by evaluating \eqref{eqn:total_range}--\eqref{eqn:targetbearing}. The $r$th column measurement vector for the hypotheses that a reflector object exist at $X$ and the null hypothesis are hence given by
\begin{equation}
\mathbf{Z}_{m}(r) =
\begin{cases}
 \alpha_{m}\mathbf{s}_{m}(r,X) + \mathbf{n}_{m}(r)
&,H_1 \, \text{holds}, \\
\mathbf{n}_{m}(r) &,H_0 \, \text{holds},  
\end{cases}
\label{eqn:measurement_hypotheses}
\end{equation}
where $\mathbf{n}_{m}(r) \sim \mathcal{CN}(.;\mathbf{0}, \Sigma_{m})$ models the noise background of the $m$th channel and is a complex Gaussian random variable with zero mean and covariance of $\Sigma_{m}$.

Note that, for the (local) mono-static channel $m=1$, and, \eqref{eqn:measurement_hypotheses} is found for $\tau_m = 2R/c$ and the synchronisation term $\Delta t_{m} = 0$ in \eqref{eqn:column_data_model}. For $m>1$, the measurement vectors are associated with the (remote) bi-static channels, and $\Delta t_{m}$ is non-zero and unknown.


\subsection{Problem definition}

We would like to perform a hypothesis test based on the measurement model in \eqref{eqn:measurement_hypotheses}. These measurements are complex numbers and we are interested in the evaluation of the sufficient statistics for the two hypothesis. Detection/processing using complex measurements are often referred to as coherent detection/processing and conventionally the input is the same resolution bin over multiple pulse returns~\cite{Richards2005b}. Therefore, in order for this operation to maintain coherence, the target position should not be changed.

In order to extend coherent processing to the case of manoeuvring objects and remote transmitters, we introduce the mathematical statement of the problem as evaluation of a likelihood-ratio i) using complex versions of measurements (as opposed to, for example, using only their moduli) for all $M$ reflection channels, and, ii) for a time window of $K$ CPIs given an object trajectory $\{X_{k}\}_{k=1}^{K}$ where $X_{k} = [x_k,y_k,\dot{x}_k,\dot{y}_k]^{T}$ is the object kinematic state at the $k$th CPI. This likelihood ratio will then be tested against a threshold in a Neyman-Pearson sense~\cite[Chp.3]{Kay1998}. The detector we consider hence takes the form
\begin{equation}
 L(\mathbf{Z}_{1,1:K},\ldots,\mathbf{Z}_{M,1:K}| X_{1:K}, \boldsymbol{\alpha}, \Delta\mathbf{t})\\
 \underset{H_{0}}{\overset{H_{1}}{\gtrless}} \mathcal{T}
 \label{eqn:lhoodratiotest}
\end{equation}
where $\mathbf{Z}_{m,1:K}$ are the data cubes for channel $m$ over $k=1,\ldots,K$. Here, $\boldsymbol{\alpha}$ and $\Delta\mathbf{t}$ are reflectivity and synchronisation vectors across the channels, respectively, defined by 
\begin{eqnarray}
 \boldsymbol{\alpha} &\triangleq & [\alpha_{1,1}, \ldots,\alpha_{1,K},\ldots,\alpha_{M,1},\ldots,\alpha_{M,K}], \nonumber \\
 \Delta \mathbf{t} & \triangleq & [\Delta t_1,\Delta t_2,\ldots,\Delta t_M]. \nonumber
\end{eqnarray}

In order to carry out the test in \eqref{eqn:lhoodratiotest}, the trajectory $X_{1:K}$ needs to be estimated. This is also referred to as tracking and is the subject of Section~\ref{sec:Simultaneous_tracking_and_time_integration} along with estimation of the reflection coefficients $\boldsymbol{\alpha}$. Algorithmic strategies for estimating the synchronisation term $\Delta \mathbf{t}$ are introduced in Section~\ref{sec:MLEstimation}. These results are combined in Section~\ref{sec:Long_time_integration} and threshold selection is detailed in order to evaluate the detection test in \eqref{eqn:lhoodratiotest}.

\subsection{Sufficient statistics for the likelihood ratio}
\label{sec:sufficient_stats}
The likelihood ratio on the left hand side of \eqref{eqn:lhoodratiotest} factorises over as the noise samples for different CPIs are also independent. Each time term also factorises over channel likelihood ratios as the related parameters are independent, i.e.,
\begin{eqnarray}
 L = \prod_{k=1}^K  \prod_{m=1}^M \frac{ l(\mathbf{Z}_{m,k}|X_k, \alpha_{m,k}, \Delta t_m, H=H_1 )}{l(\mathbf{Z}_{m,k}|X_k, \Delta t_m, H=H_0 )}.
 \label{eqn:LRT_all}
\end{eqnarray}

These measurements also satisfy a locality property in that the number of range bins which are associated with $X_k$ is limited by the support of $\Lambda$ in~\eqref{eqn:orthogonal_wave_auto} which is of duration $2T_p$. Let us define the (range) extend of an object as
\begin{equation}
 {\cal E}_m(X_k) = \begin{cases} \{r_{m,k},r_{m,k}+1\}, & r_{m,k}T_p < \tau_{m}(X_{k})+\Delta t_{m}\\ 
 \{r_{m,k}\}, & r_{m,k}T_p = \tau_{m}(X_{k})+\Delta t_{m}\\
 \{r_{m,k}-1, r_{m,k}\}, & r_{m,k}T_p > \tau_{m}(X_{k})+\Delta t_{m}
 \end{cases},
 \label{eqn:setOfrange}
\end{equation}
where
\begin{equation}
	r_{m,k} \triangleq \left[\frac{\tau_{m}(X_{k})+\Delta t_{m}}{T_p}\right],
	\label{eqn:range bin}
\end{equation}
with $[.]$ denoting the nearest integer function, and $\tau_{m}(X_{k})+\Delta t_{m}$ gives the time of flight in the $m$th channel associated with the object state $X_{k}$. This range bin has the highest signal-to-noise ratio (in the $m$th channel) given that $\Lambda$ as a time auto-correlation function typically vanishes towards tails. 

As a result, the likelihood ratio in~\eqref{eqn:LRT_all} further decomposes into factors over range bins as       
\begin{equation}
 L =\prod_{k=1}^K \prod_{m=1}^M \prod_{r \in {\cal E}_m(X_k)} \frac{l(\mathbf{Z}_{m,k}(r)|X_k, \alpha_{m,k}, \Delta t_m, H=H_1)}{l(\mathbf{Z}_{m,k}(r)| H=H_0)},
 \label{eqn:LRT}
\end{equation}

The numerator terms in \eqref{eqn:LRT} can easily be found using the distribution of the noise in the signal model in \eqref{eqn:measurement_hypotheses} as
\begin{eqnarray}
&l&(\mathbf{Z}_{m,k}(r)| X_k, \alpha_{m,k},\Delta t_{m},H=H_1) \nonumber\\ 
&&=\mathcal{CN}\Big(\mathbf{Z}_{m,k}(r); \alpha_{m,k}\mathbf{s}_{m}(r,X_{k}) , \Sigma_{m}\Big).
\label{eqn:LRT_numerator_remte}
\end{eqnarray}
The denominator in \eqref{eqn:LRT} regarding the noise only hypothesis is nothing but the noise density evaluated at $\mathbf{Z}_{m,k}(r)$. Therefore, the instantaneous likelihood ratio in \eqref{eqn:LRT} after substituting from \eqref{eqn:LRT_numerator_remte}~and~the noise distribution is found as
\begin{eqnarray}
&&\mspace{-20mu}L(\mathbf{Z}_{m,k}(r)|X_{k}, \alpha_{m,k}, \Delta t_m) \nonumber\\ 
&\triangleq&\frac{\mathcal{CN}\Big(\mathbf{Z}_{m,k}(r); \alpha_{m,k}\mathbf{s}_{m}(r,X_{k}) , \Sigma_{m}\Big)}{\mathcal{CN}\Big(\mathbf{Z}_{m,k}(r_{m,k}); \vect{0}, \Sigma_{m}\Big)} \nonumber \\
&=&\exp \Big\{ 2 \mathrm{Re}\big\{ \alpha^*_{m,k}\mathbf{s}_{m}(r,X_{k})^H \Sigma^{-1}_m \mathbf{Z}_{m,k}(r)  \big\} \Big\} \nonumber \\
&&\mspace{-30mu}\times\exp\Big\{ -|\alpha_{m,k}|^2\mathbf{s}_{m}(r,X_{k})^H \Sigma^{-1}_m \mathbf{s}_{m}(r,X_{k}) \Big\},
\label{eqn:instantaneous_LRT}
\end{eqnarray}
where $(.)^H$ is the Hermitian of its argument, $\mathrm{Re}\{.\}$ takes the real part of its complex argument, $()^*$ denotes conjugate and $|.|$ denotes modulus of a complex variable, respectively.

The likelihood ratio evaluation given in \eqref{eqn:instantaneous_LRT} is advantageous in that only a linear operation needs to be performed on the measurements which is in the form of a whitening transform with the inverse noise covariance followed by an inner product with the signal model. Because the signal model involves the spatial steering vector in~\eqref{eqn:ULA_model}, this inner product effectively performs beam-forming on the measurements filtering out contributions of other objects at the same range. Note that a second filtering is with respect to the Doppler as the temporal steering vector~\eqref{eqn:temporal_model} is also in the signal model.

\section{Simultaneous tracking and reflection coefficient estimation}
\label{sec:Simultaneous_tracking_and_time_integration}
In this section, we consider estimation of the object trajectory $X_{1:K}$ using coherent pulse returns during a CPI. Object trajectories are modelled as random vector sequences generated by a Markov state space model~\cite{B.RisticGordon2004}, i.e.,
\begin{equation}
 X_{1:K} \sim p(X_1)\prod_{k=2}^K p(X_k|X_{k-1}),
\end{equation}
where the Markov transition density is selected as 
\begin{eqnarray}
&&p(X_{k}|X_{k-1}) = \mathcal{N}(X_{k};FX_{k-1}, Q) \nonumber\\
&&\quad F = \begin{bmatrix}
1&0&\Delta&0\\
0&1&0&\Delta\\
0&0&1&0\\
0&0&0&1
\end{bmatrix},
\label{eqn:Markov_transition}
\end{eqnarray}  
where $\Delta$ is the time interval between two consecutive pulse train transmissions (or, the illumination period), $F$ models constant velocity motion, and {$Q$ is the covariance matrix specifying the level of the process noise modelling unknown manoeuvres~$\text{\cite[Chp.6]{YaakovBar-Shalom2001}}$. For example, a variance of $\sigma^{2}_{v}$ in each direction of the velocity is modelled with}
\begin{equation}
	{\quad Q = \sigma^{2}_{v}} \times\begin{bmatrix}
	\frac{\Delta^{3}}{3}&0&\frac{\Delta^{2}}{2}&0\\
	0&\frac{\Delta^{3}}{3}&0&\frac{\Delta^{2}}{2}\\
	\frac{\Delta^{2}}{2}&0&\Delta&0\\
	0&\frac{\Delta^{2}}{2}&0&\Delta
\end{bmatrix}.
\label{eqn:Q_covariance} 
\end{equation}

The initial distribution $p(X_1)$ is selected as a uniform distribution over the range-bearing interval for the detection test. These intervals often correspond to radar specific resolution bins. Let us denote the corresponding bounded set in the state space by $\cal B$, and a uniform distribution on $\cal B$  by $U_{\cal B}$. Therefore,
\begin{equation}
 p(X_1) = U_{\cal B}(X_1).
 \label{eqn:Initial_state}
\end{equation}

Sequential estimation of $X_{1:K}$ as data cubes arrive is performed by using Bayesian recursive filtering~\cite{B.RisticGordon2004}. Suppose we have the given distribution of the state variable at the time step $k-1$ based on all the measurements collected up to and including CPI $k-1$, i.e., $p(X_{k-1}|{\bf{Z}}_{1:k-1})$. In order to update this prior information with the measurement at the $k$th CPI, first, the Chapman-Kolmogorov equation is realised and a prediction density is found as 
\begin{equation}
p(X_k|\mathbf{Z}_{1:k-1})= \int p(X_k|X_{k-1})p(X_{k-1}|\mathbf{Z}_{1:k-1})\mathrm{d}X_{k-1},
\label{eqn:Bayesian_prediction}
\end{equation}
where the first term inside the integral is the Markov transition given by~\eqref{eqn:Markov_transition}.

The update stage of the filtering is given by multiplying this prediction and the measurement likelihood together with marginalising out all other variables, i.e.,
\begin{eqnarray}
p(X_k|\mathbf{Z}_{1:k}) &\propto& \int_{\boldsymbol{\alpha}_{k}}\int_{\Delta\mathbf{t}}l(\mathbf{Z}_{k}|X_k, \boldsymbol{\alpha}_{k},\Delta\mathbf{t}) \nonumber\\
&&\times p(\boldsymbol{\alpha}_{k})p(\Delta\mathbf{t})p(X_k|\mathbf{Z}_{1:k-1})\mathrm{d}\boldsymbol{\alpha}_{k}\mathrm{d}\Delta\mathbf{t},
\label{eqn:Bayesian_update}
\end{eqnarray}
where $p(\boldsymbol{\alpha}_{k})$ and $p(\Delta\mathbf{t})$ are prior densities for the reflection coefficient and the synchronisation term, respectively.

The measurement likelihood in~\eqref{eqn:Bayesian_update} is the product of the numerator terms in the likelihood ratio in~\eqref{eqn:LRT} over the object's range bins and channels for the time step $k$,~i.e.,
\begin{eqnarray}
&l&(\mathbf{Z}_{k}|X_k, \boldsymbol{\alpha}_{k}, \Delta\mathbf{t}) \nonumber\\
&&\propto \prod_{m=1}^{M}\prod_{r \in {\cal E}_m(X_k)}l(\mathbf{Z}_{m,k}(r)|X_k, \alpha_{m,k}, \Delta t_m, H=H_1),
\label{eqn:new_update1}
\end{eqnarray}
and is easily computed by evaluating complex Gaussian densities as discussed in Section~\ref{sec:sufficient_stats}.

The marginalisation of the reflection coefficients and synchronisation terms, however, is not straightforward: First, one needs to select prior densities for these terms. One reasonable approach is to use a non-informative prior such as Jeffrey's prior~\cite[Chp.5]{Murphy}. These priors are useful when they lead to tractable computations in~\eqref{eqn:Bayesian_update} (see, e.g.,~\cite{OrtonFitzgerald2002}). In our problem setting, however, Jeffrey's priors for the reflection coefficients and the synchronisation terms are constant, and, do not help in finding a tractable form for the full Bayesian update in~\eqref{eqn:Bayesian_update}.

In order to tackle this challenge, we use an empirical Bayes approach~\cite{CarlinLouis2010}. These methods approximate the integration in~\eqref{eqn:Bayesian_update} by solving an optimisation problem for finding the likely values of the unknowns and evaluating the integrand at those values. In other words, \eqref{eqn:Bayesian_update} is rewritten as
\begin{eqnarray}
 &&p(X_k|\mathbf{Z}_{1:k}) = \notag\\
 &&\int_{\boldsymbol{\alpha}_{k}}\int_{\Delta\mathbf{t}} p(X_k|\mathbf{Z}_{1:k}, \boldsymbol{\alpha}_{k},\Delta \mathbf{t} )p(\boldsymbol{\alpha}_{k},\Delta \mathbf{t}|\mathbf{Z}_{1:k}  ) \mathrm{d}\boldsymbol{\alpha}_{k}\mathrm{d}\Delta\mathbf{t}.
 \label{eqn:posterior}
\end{eqnarray}

Here, the reflection coefficients and the synchronisation terms act as model parameters to be selected and the second term inside the integration is similar to a prior for them. Because this prior is conditioned on the measurements, more probability mass should be concentrating at the maximum likelihood (ML) estimates of these values. Let us select this density as
\begin{eqnarray}
 p(\boldsymbol{\alpha}_{k},\Delta \mathbf{t}|\mathbf{Z}_{1:k}  ) &= &p(\boldsymbol{\alpha}_{k}|\mathbf{Z}_{1:k}  ) p(\Delta \mathbf{t}|\mathbf{Z}_{1:k}  ) \notag \\
 p(\boldsymbol{\alpha}_{k} |\mathbf{Z}_{1:k}  )&\leftarrow& \delta_{\hat {\boldsymbol{\alpha}
 }_k } ( \boldsymbol{\alpha}_k ) \notag \\
  p(\Delta \mathbf{t}|\mathbf{Z}_{1:k}  )&\leftarrow& \delta_{\Delta \hat{\mathbf{t}} }(\Delta \mathbf{t}), 
  \label{eqn:empiricalPriors}
\end{eqnarray}
where $\leftarrow$ denotes assignment and $\delta$ is Dirac's delta distribution. In other words, we select the model densities given the measurements as a Dirac's delta distribution concentrated {in the vicinity of their ML estimates} $\hat {\boldsymbol{\alpha}
 }_k$ and $\Delta \hat {\mathbf{t}}$, respectively.

 After substituting from the empirical priors in~\eqref{eqn:empiricalPriors} into ~\eqref{eqn:posterior}, one obtains the empirical Bayes update as
\begin{eqnarray}
p(X_k|\mathbf{Z}_{1:k}) &\approxprop& l(\mathbf{Z}_{k}|X_k, \hat{ \boldsymbol{\alpha}}_{k}, \Delta\hat{\mathbf{t}} ) p(X_k|\mathbf{Z}_{1:k-1})
\label{eqn:empiricalBayesian_update}
\end{eqnarray}
where $\approxprop$ denotes approximate proportionality. The approximation accuracy is better when these ML estimates are obtained using informative likelihoods (as quantified by their Fisher information) and equivalently have small CRLBs. 

We will detail ML estimation of the reflection coefficients~$\boldsymbol{\alpha}$ and the synchronisation terms~$\Delta\mathbf{t}$ in Section~\ref{sec:MLEstimation}. For the remaining part of this section, let us assume that these estimates are given.


For realising the recursive filtering, a sequential Monte Carlo (SMC) approach known as the particle filter is used \cite{Arulampalam2002}. In particular, we use a bootstrap filtering approach for estimating the object trajectory. 

The prediction stage at the time step $k=1$ is realised by forming a regular grid of $P$ points over $\cal B$ representing samples generated from the initial state distribution in~\eqref{eqn:Initial_state}. These points constitute an equally weighted set of particles. For $k>1$, we will have found weighted samples, or, particles, representing the state posterior in the previous step. Let us denote this set by 
$$\left\lbrace X_{k-1}^{(p)}, \zeta_{k-1}^{(p)}\right\rbrace_{p=1}^{P},$$
where $ \zeta_{k-1}^{(p)}$ is the weight of the $p$th sample. The prediction stage is then realised by sampling from the Markov transition~as
\begin{equation}
X_{k|k-1}^{(p)}~\sim~p(\,\cdot\,|X_{k-1}^{(p)}),\,\,\,p=1,\ldots,P.
\label{eqn:sampling_Markov_transition}
\end{equation}

The weights of these samples in the particle set $\{X_{k|k-1}^{(p)}, \zeta_{k|k-1}^{(p)} \}$ is given by
\begin{equation}
\zeta_{k|k-1}^{(p)} = \zeta_{k-1}^{(p)}, 
\label{eqn:pred_weights}
\end{equation}
in order for this set to represent the prediction density in~\eqref{eqn:Bayesian_prediction}.

In the update stage, the same sample set is used to represent the state posterior in~\eqref{eqn:empiricalBayesian_update}, i.e.,
\begin{equation}
 X^{(p)}_k \leftarrow X_{k|k-1}^{(p)}\,\,\,\,p=1,\ldots,P,
 \label{eqn:sampling_update}
\end{equation}
where $\leftarrow$ denotes assignment.

The weights of these samples need to be adjusted using the measurement likelihood (as per the importance sampling principle~\cite{Casella2005}), i.e.,
\begin{eqnarray}
\zeta_{k}^{(p)} &=& \frac{\tilde{\zeta}_{k}^{(p)}}{\sum_{p'=1}^{P}\tilde{\zeta}_{k}^{(p')}}, 
\label{eqn:weight_update}\\
\tilde{\zeta}_{k}^{(p)} &=&\zeta_{k|k-1}^{(p)}l(\mathbf{Z}_{k}|X_{k} = X^{(p)}_{k},\boldsymbol{\hat{\alpha}}_{k},\Delta\mathbf{\hat{t}}), \notag
\end{eqnarray}

After finding the normalised weights in~\eqref{eqn:weight_update}, we test degeneracy of the weighted particles. The degeneracy test is performed by finding the number of effective particles using
\begin{equation}
N_{eff} = \frac{1}{\sum_{p=1}^{P}\left(\zeta_{k}^{(p)}\right)^{2}},
\label{eqn:N_eff} 
\end{equation}            
and, comparing it with a threshold $\mathcal{T}_{eff}$. When $N_{eff} < \mathcal{T}_{eff}$, we perform re-sampling (see, e.g., \cite{Arulampalam2002}) and continue filtering with a new, equally weighted sample set 
$$\{ \zeta_{k}^{(p)} \leftarrow 1/P  ,X^{(p)}_k \leftarrow \tilde X^{(p)}_k\}_{p=1}^P,$$
where $\{\tilde X^{(p)} \}$ is output by the re-sampler.

Using the above particle filter, the object state $X_k$ at the $k$th CPI is estimated by using the empirical weighted average
\begin{equation}
\hat{X}_{k} = \sum_{p=1}^{P}\zeta_{k}^{(p)}X_{k|k-1}^{(p)},
\label{eqn:state_estimation}
\end{equation}
where $\hat{X}_{k}$ denotes the estimated object state $X_k$. 

A remarkable feature of the processing scheme driven by the Bayesian recursions above is that no fixed selection of the spatio-temporal steering vectors are used. The evaluation of the likelihood in the update stage in~\eqref{eqn:weight_update} specifies the steering vectors through \eqref{eqn:LRT_numerator_remte}~and~\eqref{eqn:column_data_model} as a function of the state value $X^{(p)}_{k}$. Because $X^{(p)}_{k}$ are generated by sequential processing of the data cubes over CPIs, the resulting set of spatio-temporal steering vectors adapt to the measurements. This is in stark contrast with conventional processing chains in which the bearing and Doppler space is sampled with equal size steps leading to a fixed set of steering vectors and corresponding resolution bins. Thus, a super-resolution effect is achieved when finding the object locations as demonstrated in Section~\ref{sec:Example}. 

\section{Maximum Likelihood estimation of unknown parameters}
\label{sec:MLEstimation}  
In this section, we first introduce the ML estimator for the reflection coefficients. This estimator is an iterative algorithm realising Expectation Maximisation at each step of the recursive filtering detailed in Section~\ref{sec:Simultaneous_tracking_and_time_integration}-- in particular when evaluating the tracking update in~\eqref{eqn:weight_update}--and, is also central to long time integration detailed later in Section~\ref{sec:Long_time_integration}. Second, we derive the ML synchronisation term estimator used together with the reflectivity estimator in the filter update in Section~\ref{sec:Simultaneous_tracking_and_time_integration}.
 
\subsection{ML estimation of the reflection coefficients}
\label{sec:EstimateRefCoeff}

\begin{algorithm}[t]
\caption{Particle EM algorithm for estimation of the reflection coefficients} 
\label{Algorithm:EM_algorithm}
\begin{algorithmic}[1]
\State Input: $\boldsymbol{\alpha}^{(0)}$, $\epsilon$ \Comment{Initial guess and termination threshold}
\State Input: $\{ \zeta_{k|k-1}^{(p)} X_{k|k-1}^{(p)} \}_{p=1}^P $\Comment{Particles from $p(X_k|{\mathbf{Z}_{1:k-1}})$}
\State $i \leftarrow 1$, $\boldsymbol{\alpha}^{(1)} \leftarrow \infty$ \Comment{Initialisation for the iterations}
\While{$\lVert~\boldsymbol{\alpha}^{(i)}~-~\boldsymbol{\alpha}^{(i-1)}~\rVert~>~\epsilon$}\Comment{Test convergence}
\State Find $\hat Q(\boldsymbol{\alpha}_{k},\boldsymbol{\alpha}^{(i-1)}_k)$ in~\eqref{eqn:log_likelihood_expansion} using~\eqref{eqn:Q_function_MC},~\eqref{eqn:EM_weight} \Comment{{\bf E step}}
\State Find $\boldsymbol{\alpha}^{(i)} \longleftarrow \{ \hat \alpha_{m,k} \}_{m=1}^M$ using \eqref{eqn:reflection_coefficient_estimation},\eqref{eqn:EM_weight} \Comment{{\bf M step}}
\State $i \longleftarrow i+1$  
\EndWhile
\State Return $\boldsymbol{\hat{\alpha}}_{k} \leftarrow \boldsymbol{\alpha}^{(i)}$
\end{algorithmic}
\end{algorithm}
The reflection coefficients associated with an object at state $X_k$ are unknown constants for the duration of a CPI and vary across consecutive CPIs due to the change of the object position, orientation etc. The likelihood of these reflectivities is found by multiplying the likelihood in~\eqref{eqn:new_update1} with the priors for the other parameters and marginalising them out. Let us use the empirical prior for the synchronisation term (see, e.g.,~\eqref{eqn:empiricalPriors}) obtained using the ML estimator detailed in the next section. The likelihood to be maximised for estimation is hence found as
\begin{equation}
l(\mathbf{Z}_k|\boldsymbol{\alpha}_k) =
\int_{X} l(\mathbf{Z}_k|X_k, \boldsymbol{\alpha}_k, \Delta \mathbf{t} = \Delta \hat{\mathbf{t}} )p(X_k|\mathbf{Z}_{1:k-1} )\mathrm{d}X_k.
\label{eqn:marginal_likelihood}   
\end{equation}

It is not straightforward to optimise this function due to the marginalisation involved.  
In ML problems involving such latent variables as the state variable $X_k$, the expectation maximisation (EM) method offers an iterative and gradient-free solution~\cite{Moon1996a}. In this method, starting from an initial parameter configuration $\boldsymbol{\alpha}^{(0)}_k$, an expectation that replaces the original likelihood is maximised. For the problem at hand, these iterations are given for $i=1,2,\ldots$ by
\begin{eqnarray}
 \boldsymbol{\alpha}^{(i)}_k &=& \arg \max_{\boldsymbol{\alpha}} Q(\boldsymbol{\alpha}_k,\boldsymbol{\alpha}^{(i-1)}_k  ) \notag \\
  Q(\boldsymbol{\alpha}_k,\boldsymbol{\alpha}^{(i-1)}_k  ) &\triangleq& \mathrm{E}\{  l(\mathbf{Z}_k|X_k, \boldsymbol{\alpha}_k, \Delta \mathbf{t} = \Delta \hat{\mathbf{t}} ) | \mathbf{Z}_{1:k-1}, \boldsymbol{\alpha}_k^{(i-1)}   \} \notag \\
&\propto&  \int_{X_{k}}\log l(\mathbf{Z}_{k}|X_{k},\boldsymbol{\alpha}_{k}, \Delta \mathbf{\hat{t}}) \notag \\
&& \mspace{40mu} \times p(X_{k}|\mathbf{Z}_{1:k}, \boldsymbol{\alpha}^{(i-1)}_k,\Delta \mathbf{\hat{t}}) \mathrm{d}X_{k},
\label{eqn:expectation_maximisation}
\end{eqnarray}
where $\mathrm{E}\{.\}$ denotes the expectation.
 
Let us focus on the computation of the expectation in~\eqref{eqn:expectation_maximisation} and its maximisation. The state density function underlying the expectation is a state posterior conditioned on the previously found value of the reflectivities, i.e.,
\begin{eqnarray}
 &p&(X_{k}|\mathbf{Z}_{1:k}, \boldsymbol{\alpha}_k^{(i-1)},\Delta \mathbf{\hat{t}})  \notag\\ 
 &&\propto \,\,l(\mathbf{Z}_{k}|X_{k}, \boldsymbol{\alpha}_k^{(i-1)},\Delta \mathbf{\hat{t}}) p(X_{k}|\mathbf{Z}_{1:k-1}),
\label{eqn:expectation_density}
\end{eqnarray}
where the density function on the right hand side is nothing but the predictive density of the Bayesian filtering recursions given in Section~\ref{sec:Simultaneous_tracking_and_time_integration}. Thus, the samples generated in the prediction stage in~\eqref{eqn:sampling_Markov_transition}~and~\eqref{eqn:pred_weights} lead to an importance sampling~\cite{Casella2005} estimate of the expectation. Given $\left\lbrace X_{k|k-1}^{(p)}, \zeta_{k|k-1}^{(p)}\right\rbrace_{p=1}^{P}$, this importance sampling estimate is given by
\begin{eqnarray}
&&\mspace{-80mu} \hat Q(\boldsymbol{\alpha}_{k},\boldsymbol{\alpha}^{(i-1)}_k) \notag \\
&\approxprop& \sum_{p=1}^{P} \xi^{(i-1)}_{p} \log l(\mathbf{Z}_{k}|X_{k} = X^{(p)}_{k|k-1},\boldsymbol{\alpha}_{k}, \Delta \mathbf{\hat{t}}),
\label{eqn:Q_function_MC} \\
\xi^{(i-1)}_{p}&=&\frac{l(\mathbf{Z}_{k}|X_{k}=X_{k|k-1}^{(p)}, \boldsymbol{\alpha}^{(i-1)}_k,\Delta \mathbf{\hat{t}})\zeta_{k|k-1}^{(p)}}{\sum_{p'=1}^{P}l(\mathbf{Z}_{k}|X_{k}=X_{k|k-1}^{(p')}, \boldsymbol{\alpha}^{(i-1)}_k,\Delta \mathbf{\hat{t}})\zeta_{k|k-1}^{(p')}}, 
\label{eqn:EM_weight}
\end{eqnarray}     
where $\hat Q$ denotes the estimate of the term proportional to $Q$ in~\eqref{eqn:expectation_maximisation}.

This approximation is a sum of terms quadratic in $\boldsymbol{\alpha}_{k}$. This can easily be seen by substituting from \eqref{eqn:new_update1}~and~\eqref{eqn:LRT_numerator_remte} to~\eqref{eqn:Q_function_MC}. The resulting expression is given in \eqref{eqn:log_likelihood_expansion} (see the top of the next page). After taking the first order partial derivative of this expression with respect to $\alpha_{m,k}$ and setting it to zero, the the ML estimate of the $m$th reflection channel is found in closed form given in \eqref{eqn:reflection_coefficient_estimation} (see the top of the next page).

Note that the ML estimator in \eqref{eqn:reflection_coefficient_estimation} takes the inner product of the ``whitened'' measurements with the signal model $\mathbf{s}_{m}$ given in \eqref{eqn:column_data_model} for each state particle $X_{k|k-1}^{(p)}$. This operation effectively performs digital beam-forming towards the particle state in space, and, matches its approach speed through its Doppler frequency encoded in $\mathbf{s}_{m}$. As a results, the estimator will not be interference with other objects unless they appear very close to the state value in terms of the achievable spatial and Doppler resolution.

After finding $\boldsymbol{\hat{\alpha}}_{k}^{(i)} = \{\hat{\alpha}_{m,k}\}^{M}_{m=1}$ for $M$ reflection coefficients using \eqref{eqn:reflection_coefficient_estimation}, convergence is tested by comparing the norm of the difference between the parameter configurations found in consecutive time steps with a threshold, i.e., iterations are terminated at $i$ if
$$\lVert~\boldsymbol{\alpha}^{(i)}_k~-~\boldsymbol{\alpha}^{(i-1)}_k~\rVert < \epsilon, $$
where $\lVert.\rVert$ denotes the complex Euclidean norm. A pseudo-code of these steps are given in Algorithm~\ref{Algorithm:EM_algorithm}.

\begin{figure*}[t]
	\vspace{-1.0em}
	\normalsize
	\setcounter{equation}{44}
	\begin{eqnarray}
	&&\hat Q(\boldsymbol{\alpha}_{k},\boldsymbol{\alpha}^{(i-1)})
	=\sum_{p=1}^{P}\xi^{(i-1)}_{p}\Big[\sum_{m=1}^{M}\sum_{r\in {\cal E}_m(X^{(p)}_{k|k-1})}\Big(-\log\left(\pi^{LN}\det(\Sigma_m)\right) \nonumber -\mathbf{Z}_{m,k}(r)^{H}\Sigma_m^{-1}\mathbf{Z}_{m,k}(r) \nonumber\\ &&+2\operatorname{Re}\lbrace \alpha^{*}_{m,k}\mathbf{s}_{m}(r,X^{(p)}_{k|k-1})^{H}\Sigma_m^{-1}\mathbf{Z}_{m,k}(r)\rbrace-|\alpha_{m,k}|^{2}\mathbf{s}_{m}(r,X^{(p)}_{k|k-1})^{H}\Sigma_m^{-1}\mathbf{s}_{m}(r,X^{(p)}_{k|k-1})\Big)\Big] \label{eqn:log_likelihood_expansion} \\
	&&\hat{\alpha}_{m,k}= \frac{\sum_{p=1}^{P}\sum_{r\in {\cal E}_m(X^{(p)}_{k|k-1})}\xi^{(i-1)}_{p}\mathbf{s}_{m}(r,X^{(p)}_{k|k-1})^{H}\Sigma_m^{-1}\mathbf{Z}_{m,k}(r)}{\sum_{p=1}^{P}\sum_{r\in {\cal E}_m(X^{(p)}_{k|k-1})}\xi^{(i-1)}_{p}\mathbf{s}_{m}(r,X^{(p)}_{k|k-1})^{H}\Sigma_m^{-1}\mathbf{s}_{m}(r,X^{(p)}_{k|k-1})}
	\label{eqn:reflection_coefficient_estimation}  
	\end{eqnarray}
	\hrulefill
	\vspace{-1.2em}
\end{figure*}
\subsection{Synchronisation of the local processor with remote transmitters} 
\label{sec:bistaticsynch}
In this section, we detail the ML estimation of the unknown synchronisation term $\Delta t_{m}$ parametrising the time origin shift between the local receiver and the $m$th separately located transmitter. Our approach exploits the fact the data cube for the $m$th bi-static channel contains direct path signals from the transmitter that can be recovered by diverting a digital beam towards the transmitters spatial state simultaneously with other processing tasks on the data cube, e.g., those related to trajectory estimation and reflectivity estimation for other spatio-temporal points.

The direct path signal in the $m$th channel can easily be modelled using the spatial and temporal steering vectors  defined in~Section~\ref{sec:problem_statement} in \eqref{eqn:ULA_model}~and~\eqref{eqn:temporal_model}, respectively. The state of the $m$th transmitter is given by $X_{m} = [x_m,y_m, 0, 0]^{T}$ which is associated with the time-of-flight $\tau(X_m)$ given in \eqref{eqn:tof} as the time to receiver. The angle of arrival is denoted by $\theta_{m}(X_m)$ which is computed using \eqref{eqn:targetbearing}. Different from a reflection channel, the unknown reflectivity is replaced with a known pulse energy term. Thus, the CPI measurement vector at the $r$th range bin obtained by sampling the $m$th matched filter output is given by 
\begin{eqnarray}
\mathbf{Z}_{m}(r) &=& \sqrt{E_m}\mathbf{\tilde{s}}_{m}(r, X_m) + \mathbf{n}_{m}(r), 
\label{eqn:direct channel}	   \\
\mathbf{\tilde{s}_{m}}(r, X_m) &\triangleq& s(\Delta t_{m})\mathbf{s}_{s}(\theta_{m}(X_m))\otimes\mathbf{s}_{t}(\tau(X_m), \Omega_m(X_m) = 0) \notag \\
	&& \mspace{40mu} \times\Lambda_{m}(rT_p -\tau(X_m)-\Delta t_{m}), \notag \\
	&=& s(\Delta t_{m}) s(\tau(X_m) )\times\mathbf{s}_{s}(\theta_{m}(X_m))\otimes{\mathbf{1}} \notag \\
	&& \mspace{40mu} \times\Lambda_{m}(rT_p -\tau(X_m)-\Delta t_{m}), \notag 
\end{eqnarray}
where $E_m$ is the pulse energy, $\mathbf{\tilde{s}}_{m}$ is the noise free signal model associated with the transmitter state $X_m$, and, $\mathbf{1}$ is an $N\times1$ all ones vector\footnote{Note that $\mathbf{\tilde{s}}_{m}$ differs from $\mathbf{{s}}_{m}$ in \eqref{eqn:column_data_model} in that the latter uses the bi-static time-of-flight in both the temporal steering vector and the waveform auto-correlation delay, whereas, the former uses direct path time-of-flight. Because the transmitters are of zero Doppler frequency, the temporal steering vector reduces to an all ones vector scaled with $s(\tau(X_m))$.}.

In the presence of reflectors, we will have received a superposition of this signal and reflections from different spatio-temporal states. In order to recover the direct path signal, a spatio-temporal steering vector that matches $\mathbf{\tilde{s}}_{m}$ in \eqref{eqn:direct channel} is used which is given by
\begin{eqnarray}
\mathbf{h}(X_m) &\triangleq& \mathbf{s}_{s}(\theta_{m}(X_m))\otimes\mathbf{s}_{t}(\tau(X_m), \Omega_m(X_m) = 0), \notag \\
&=&s(\tau(X_m) )\times\mathbf{s}_{s}(\theta_{m}(X_m))\otimes{\mathbf{1}} \notag.
\end{eqnarray}
Note that this filter is nothing but a (scaled) beam-forming vector diverting a beam towards $\theta_m(X_m)$ and maps the ${LN \times 1}$ measurement vector $\mathbf{Z}_{m}(r)$ to a single complex value given by
\begin{eqnarray}
d_m(r) &\triangleq& \mathbf{h}(X_m)^H\mathbf{Z}_{m}(r) \nonumber\\
&=&LN\sqrt{E_m}s(\Delta t_{m})\Lambda_{m}(rT_p -\tau(X_m)-\Delta t_{m}) \notag\\
&&+ n_{m}(r).
\label{eqn:direct_Ch}
\end{eqnarray}
Here, the noise term is the inner product of the beam-forming vector and the complex Gaussian measurement noise in~\eqref{eqn:measurement_hypotheses},~i.e., 
$$n_{m}(r) = \mathbf{h}(X_m)^H\mathbf{n}_{m}(r)$$
which itself is a random variable with a complex Gaussian distribution of mean zero and variance $\sigma^{2}_{d,m} = \mathbf{h}(X_m)^{H}\Sigma_{m} \mathbf{h}(X_m)$.

As a result, the likelihood to be maximised is 
\begin{equation}
 l( d_m(1),\ldots,d_m(\Gamma) | \Delta t ) = \prod_{r=1}^\Gamma {\cal CN}( d_m(r);  \mu_{d,m}(\Delta t),\sigma^{2}_{d,m})
 \end{equation}
 where the expected value of the complex Gaussian distributions as a function of $\Delta t$ is given by
 \begin{eqnarray}
 \mu_{d,m}(\Delta t) &=& \sqrt{E_m}LN \exp(-j\omega_c \Delta t_m)\notag\\ 
 &&\times\Lambda_{m}(rT_p -\tau(X_m)-\Delta t_{m}).
 \label{eqn:mu_deltat}
\end{eqnarray}

Here, only those range bins for which the argument of $\Lambda_m$ falls within $(0,2T_p)$ contribute to the maximisation -- otherwise, the corresponding distribution is same with that for the noise term.These range bins are given by
\begin{equation}
	{\cal \tilde{E}}_m(\Delta t) = \begin{cases} \{r_{m},r_{m}+1\} & r_{m}T_{p} < \tau(X_m)+\Delta t_m \\
	\{r_{m}\} & r_{m}T_{p} = \tau(X_m)+\Delta t_m \\
	\{r_{m},r_{m}-1\} & r_{m}T_{p} > \tau(X_m)+\Delta t_m 
	\end{cases},
	\label{eqn:range_of_direct_channel}
\end{equation}
where
\begin{equation*}
	r_{m} = \left[ \frac{\tau(X_m)+\Delta t_m}{T_p}\right].
\end{equation*}

Thus, the ML estimator that takes into account $k$ data cubes at time $k$ starting from the first one is given by
\begin{eqnarray}
 \Delta \hat{t}_{m}&=& \arg \max_{\Delta t_m} J_k(\Delta t_m) \notag \\
 J_k(\Delta t_m) &=& \log \prod_{k'=1}^{k}\prod_{r \in {\cal \tilde{E}}_m(\Delta t)} \mspace{-20mu}{\cal CN}(d_{m,k'}(r); \mu_{d,m}(\Delta t), \sigma^2_{d,m} ) \notag \\
 &\propto& \sum_{k'=1}^{k} \sum_{r \in  {\cal \tilde{E}}_m(\Delta t) }  \left( d_{m,k'}(r) - \mu_{d,m}(\Delta t)  \right)^* \notag \\
 &&\mspace{100mu} \times \left( d_{m,k'}(r) - \mu_{d,m}(\Delta t)  \right).
 \label{eqn:ObjectiveDeltat}
\end{eqnarray}

Here, the relation between $\Delta t$ and the objective function $J_k$ is a concave relation on the average (and as $k$ increases, asymptotically). However,~\eqref{eqn:mu_deltat} does not yield a closed form solution and render gradient-free iterative methods such as one-dimensional line search techniques~\cite{Bazaraa1993} as better alternatives. These algorithms require only evaluation of~\eqref{eqn:ObjectiveDeltat} and iteratively reduce an initially selected interval of uncertainty.

We use the golden section search algorithm~\cite{Bazaraa1993} and select and initial interval for $\Delta t$ based on a preliminary search over the grid of values $\Delta t \in \{0,T_p,2T_p,\ldots,(\Gamma-1)T_p \}$ which yields a rough estimate. Let us denote this term by $\Delta \hat t_0$~\footnote{Note that, equivalently found is $$\hat r_m = \left[ \frac{\tau(X_m)+ \Delta \hat t_0 }{T_p}\right].$$}. The initial interval of uncertainty is selected as $[\Delta \hat t_0 - T_p, \Delta \hat t_0+T_p]$. The golden section search reduces the width of this interval exponentially to a ratio of $(0.618)^{\nu-1}$ after $\nu$ iterations~\cite{Bazaraa1993}. Therefore, in eight iterations, this width reduces below one tenth of a pulse duration, i.e., $T_p/10$. This search is detailed in~Algorithm~\ref{Algorithm:Golden_section_search_1}.               
 
\begin{algorithm}[t]
\caption{Maximum likelihood estimation of $\Delta t_{m}$ via golden section line search: The initial interval of uncertainty is selected as $[\Delta \hat t_0 - T_p, \Delta \hat t_0 + T_p]$ as detailed in Section~\ref{sec:bistaticsynch}.} 
\label{Algorithm:Golden_section_search_1}
\begin{algorithmic}[1]
\State {Input:} $[\Delta t_1,\Delta t_2]$, $\epsilon$ \Comment{Initial interval of uncertainty and termination threshold}
\State $\alpha \leftarrow 0.618$
\State $\Delta \tilde t_1 \leftarrow \Delta t_1 + (1-\alpha)(\Delta t_2 - \Delta t_1)$ \Comment{Evaluation point 1} 
\State $\Delta \tilde t_2 \leftarrow \Delta t_1 + \alpha(\Delta t_2 - \Delta t_1)$ \Comment{Evaluation point 2}
\State Compute $J_k(\Delta \tilde t_1)$ and $J_k(\Delta \tilde t_2)$ using~\eqref{eqn:ObjectiveDeltat},~\eqref{eqn:mu_deltat},~\eqref{eqn:direct_Ch}
\While{$|\Delta t_2-\Delta t_1| > \epsilon$} \Comment{Until $\epsilon$ accuracy is reached}
\If{$J_k(\Delta \tilde t_1)>J_k(\Delta \tilde t_2)$} 
\State $\Delta t_2 \leftarrow \Delta \tilde t_2$ \Comment{New interval:$[\Delta t_1, \Delta \tilde t_2]$}
\State $\Delta \tilde t_2 \leftarrow \Delta \tilde t_1$, $J_k(\Delta \tilde t_2) \leftarrow J_k(\Delta \tilde t_1)$ \Comment{Assignments}
\State $\Delta \tilde t_1 \leftarrow \Delta t_1+(1-\alpha)(\Delta t_2-\Delta t_1)$ 
\State Compute $J_k(\Delta \tilde t_1)$ using~\eqref{eqn:ObjectiveDeltat},~\eqref{eqn:mu_deltat},~\eqref{eqn:direct_Ch}\Comment{New evaluation} 
\Else
\State $\Delta t_1 \leftarrow \Delta \tilde t_1$ \Comment{New interval:$[\Delta \tilde t_1, \Delta t_2]$}
\State $\Delta \tilde t_1 \leftarrow \Delta \tilde t_2$, $J_k(\Delta \tilde t_1)\leftarrow J_k(\Delta \tilde t_2)$ \Comment{Assignments}
\State $\Delta \tilde t_2 \leftarrow \Delta t_1+\alpha(\Delta t_2-\Delta t_1) $
\State Compute $J_k(\Delta \tilde t_2)$ using~\eqref{eqn:ObjectiveDeltat},~\eqref{eqn:mu_deltat},~\eqref{eqn:direct_Ch}\Comment{New evaluation} 
\EndIf
\EndWhile
\If{$J_k(\Delta \tilde t_1)>J_k(\Delta \tilde t_2)$}
\State Return $\Delta\hat{t}_m = \Delta t_1$
\Else
\State Return $\Delta\hat{t}_m = \Delta t_2$
\EndIf
\end{algorithmic}
\end{algorithm}

\section{Long time integration for detection}
\label{sec:Long_time_integration}
In this section, we detail the evaluation of the statistical test given in~\eqref{eqn:lhoodratiotest}. The sufficient statistics of this test is given in Section~\ref{sec:sufficient_stats}, in particular in~\eqref{eqn:LRT}--\eqref{eqn:instantaneous_LRT}. Here, we first combine the results from Sections~\ref{sec:Simultaneous_tracking_and_time_integration}~and~\ref{sec:MLEstimation} into a single algorithm. Then, in Section~\ref{sec:CFARThreshold}, we provide explicit formulae for finding the threshold as a function of a selected constant false alarm rate $P_{fa}$ and integration time $k$.

In order to evaluate the likelihood ratio in~\eqref{eqn:LRT}, we first  estimate $\Delta\mathbf{t}$ using~Algorithm~\ref{Algorithm:Golden_section_search_1} for each bi-static channel. Given this quantity, we sequentially estimate the target trajectory $\hat X_{k}$ (Section~\ref{sec:Simultaneous_tracking_and_time_integration} ) and the reflection coefficients $\hat{\boldsymbol{\alpha}}_{k}$ using the EM iterations in Algorithm~\ref{Algorithm:EM_algorithm} over $k=1,\ldots,K$. As such, the integration of instantaneous likelihood ratios in \eqref{eqn:instantaneous_LRT} -- given the aforementioned estimates-- into the test value in \eqref{eqn:LRT} is carried out recursively. For this purpose, let us define the logarithm of the test value at $k$ as
\begin{eqnarray}
 \log L_k &\triangleq& \sum_{k'=1}^k \sum_{m=1}^M \sum_{r \in {\cal E}_m(\hat X_k)} L(\mathbf{Z}_{m,k'}(r)|\hat X_{k'}, \hat \alpha_{m,k'}, \Delta \hat t_m) \notag \\
 &=& \log L_{k-1} + L(\mathbf{Z}_{k}(\hat X_k)|\hat X_{k}, \hat {\boldsymbol{\alpha}}_{k}, \Delta \hat {\mathbf t})
 \label{eqn:integration}
\end{eqnarray}
where
\begin{eqnarray}
 && \mspace{-20mu} L(\mathbf{Z}_{k}(\hat X_k)|\hat X_{k}, \hat {\boldsymbol{\alpha}}_{k}, \Delta \hat {\mathbf t}) \notag \\
  &\triangleq& \sum_{m=1}^M \sum_{r \in {\cal E}_m(\hat X_k)} L(\mathbf{Z}_{m,k'}(r)|\hat X_{k'}, \hat \alpha_{m,k'}, \Delta \hat t_m)\notag \\
  &=& \sum_{m=1}^M\sum_{r \in {\cal E}_m(\hat{X}_{k})} \Big(2 \mathrm{Re}\big\{ \hat{\alpha}^*_{m,k}\mathbf{s}_{m}(r,\hat{X}_{k})^H \Sigma^{-1}_m \mathbf{Z}_{m,k}(r)\big\} \Big. \notag \\
&&\Big. -|\hat{\alpha}_{m,k}|^2\mathbf{s}_{m}(r,\hat{X}_{k})^H \Sigma^{-1}_m \mathbf{s}_{m}(r,\hat{X}_{k})\Big).
\label{eqn:InslogL}
\end{eqnarray}
Here, \eqref{eqn:InslogL} is the contribution of the measurements at time $k$ into the integration in~\eqref{eqn:integration}. The proposed processing performs coherent integration of ${\cal E}_m(\hat{X}_{k}) \times L \times N$ samples during a CPI in each channel. The integration is non-coherent across the channels as well as consecutive CPIs. The key is that the object trajectory is taken into account when performing all these simultaneously.

The object detection is hence performed by comparing the output of the aforementioned log-likelihood ratio to a detection threshold, i.e.,
\begin{equation}
	\log L_K \underset{H_{0}}{\overset{H_{1}}{\gtrless}} \log \mathcal{T}_{K},
	\label{eqn:Detection_scheme}
\end{equation}
where $\log \mathcal{T}_{K}$ is the detection threshold for a given constant false alarm rate (CFAR) for $K$ steps of integration. The next section details the computation of this threshold value. A pseudo-code of the overall process is given in~Algorithm~\ref{Algorithm:proposed_long_time_algorithm}.  

\begin{algorithm}[t]
\caption{The proposed simultaneous tracking and long time integration algorithm} 
\label{Algorithm:proposed_long_time_algorithm}
\begin{algorithmic}[1]
\State {Input:} Data cubes $\mathbf{Z}_{m,k}$ for channels $m=1,\ldots,M$, time steps $k=1,\ldots,K$ \Comment{ see~\eqref{eqn:measurement_hypotheses}  }
\State {Input:} Detection threshold ${\cal T}_K$
\State {Initialisation: Generate particles in the cell under test} $\left\lbrace X_{1}^{(p)}, \zeta_{1}^{(p)}\right\rbrace_{p=1}^{P}$ \Comment{see~\eqref{eqn:Initial_state}}
\State {Initialisation: $\log L_0 \leftarrow 0$}
\For{$k = 1,\ldots,K$}
\If{$k\geq2$}\Comment{Prediction stage}
\State Generate $\left\lbrace X_{k|k-1}^{p}, \zeta_{k|k-1}^{p}\right\rbrace_{p=1}^{P}$ \Comment{see~\eqref{eqn:sampling_Markov_transition},~\eqref{eqn:pred_weights}}
\EndIf
\State $\text{Find}$ $\Delta\hat{\mathbf{t}}$ using Algorithm~\ref{Algorithm:Golden_section_search_1} for $m=2,\ldots,M$ \Comment{see Section~\ref{sec:bistaticsynch}}
\State Find $\boldsymbol{\hat{\alpha}}_k$ using the EM iterations in Algorithm~\ref{Algorithm:EM_algorithm}
\State Update $\{ X_k^{(p)},\zeta_{k}^{(p)}\}_{p=1}^P $ using~\eqref{eqn:sampling_update}, \eqref{eqn:weight_update}
\Comment{Update stage}
\State Estimate $\hat{X}_k$ using \eqref{eqn:state_estimation}
\State Compute $ L(\mathbf{Z}_{k}(\hat X_k)|\hat X_{k}, \hat {\boldsymbol{\alpha}}_{k}, \Delta \hat {\mathbf t})$ using~\eqref{eqn:InslogL}
\State $\log L_k = \log L_{k-1} + L(\mathbf{Z}_{k}(\hat X_k)|\hat X_{k}, \hat {\boldsymbol{\alpha}}_{k}, \Delta \hat {\mathbf t})$ \Comment{Integration step, see \eqref{eqn:integration}}
\EndFor
\If{$\log L_K > \log \mathcal{T}_{K}$} \Comment{The detection test in \eqref{eqn:Detection_scheme}}
\State Return $H_1$
\Else 
\State Return $H_0$
\EndIf
\end{algorithmic}
\end{algorithm}    

\subsection{Constant false alarm rate threshold for the detection test}
\label{sec:CFARThreshold}
In the hypothesis test in~\eqref{eqn:Detection_scheme} it is highly desirable to select a threshold $\mathcal{T}_{K}$ that yields a selected constant false alarm rate (CFAR) $P_{fa}$. For the calculation of $\mathcal{T}_{K}$ as function of $P_{fa}$, we consider the distribution of the likelihood ratio given in~\eqref{eqn:LRT} under the $H=H_0$ hypothesis for the measurement in~\eqref{eqn:measurement_hypotheses}~\cite{Kay1998}. Let us find the logarithm of the likelihood ratio after substituting from \eqref{eqn:instantaneous_LRT} into \eqref{eqn:LRT}:
\begin{equation}
 \eta_K \triangleq \log L_K = \sum_{k=1}^K \sum_{m=1}^M \sum_{r \in {\cal E}_m(X_k)} \eta_{m,k,r}
 \label{eqn:Globaleta}
\end{equation}
where the terms inside the summations are given by
\begin{eqnarray}
\eta_{m,k,r}&= &2 \mathrm{Re}\{ \mathbf{s}_{m,k,r}^H  \Sigma^{-1}_m \mathbf{Z}_{m,k}(r) \} - \mathbf{s}_{m,k,r}^H  \Sigma^{-1}_m \mathbf{s}_{m,k,r} \notag \\ 
 \mathbf{s}_{m,k,r} &=& \alpha_{m,k}\mathbf{s}_{m}(r,X_{k}).
 \label{eqn:knownsignal}
\end{eqnarray}

The distribution of the real variable $\eta_{m,k,r}$ is a Gaussian when the signal model $\mathbf{s}_{m,k,r}$ is known and the measurements $\mathbf{Z}_{m,k}(r)$ are generated from a complex Gaussian~\cite[Chp.13]{Kay1998},~i.e., $\eta_{m,k,r} \sim {\cal N}(.;\mu_{m,k,r}, \sigma^2_{m,k,r} )$, with the moments given by
\begin{eqnarray}
 \mu_{m,k,r} = - \mathbf{s}_{m,k,r}^H  \Sigma^{-1}_m \mathbf{s}_{m,k,r} ,
 \notag \\ 
 \sigma^2_{m,k,r} = 2 \mathbf{s}_{m,k,r}^H  \Sigma^{-1}_m \mathbf{s}_{m,k,r}.
 \notag 
\end{eqnarray}
Owing to the independence of the noise samples, $\eta_K$ is also Gaussian for the case, i.e., $\eta_K \sim {\cal N}(.;\mu_K,\sigma^2_K)$, with the moments given by
\begin{eqnarray}
\mu_{K} &= &\sum_{k=1}^K \sum_{m=1}^M \sum_{r \in {\cal E}_m(X_k)} \mu_{m,k,r} \\
\sigma^2_{K} &=& \sum_{k=1}^K \sum_{m=1}^M \sum_{r \in {\cal E}_m(X_k)} \sigma^2_{m,k,r}.
\end{eqnarray}

Therefore, the probability of false alarm $P_{fa}$ is related to the test variable $\eta_K$~in~\eqref{eqn:Globaleta} and the threshold ${\cal T}_K$ through
\begin{eqnarray}
 P_{fa} &= &Pr\{ \eta_K > \log {\cal T}_K | H=H_0 \}  \nonumber \\ 
&=& \int_{\log \mathcal{T}_K}^{+\infty}{\cal N}(\eta'_K;\mu_K,\sigma^2_K)\mathrm{d}\eta'_K\nonumber\\ 
&=& Q\left( \frac{ \log \mathcal{T}_K - \mu_K }{\sigma_K} \right) \nonumber
\end{eqnarray}
where $Q(.)$ denotes the tail probability function of the standard normal distribution~\cite{Kay1998}. As a result, the threshold $\mathcal{T}_K$ given $P_{fa}$ for $K$ steps of integration is found as
\begin{equation}
\mathcal{T}_K = \exp\left( Q^{-1}( P_{fa})\sigma_K + \mu_K   \right).
\label{eqn:CFAR_Threshold}
\end{equation}

As a summary, a CFAR threshold for the proposed long time integration method is calculated using \eqref{eqn:knownsignal}--\eqref{eqn:CFAR_Threshold} given the true values of the reflectivities and the object trajectory specifying~\eqref{eqn:knownsignal}. This clairvoyant threshold is used in the next section, for comparing Algorithm~\ref{Algorithm:proposed_long_time_algorithm} with the clairvoyant integrator and a conventional alternative.

\subsection{{Signal to noise ratio (SNR) in the radar data cube}}
\label{sec:SNR}
{Here, we provide explicit formulae for the signal to noise ratio (SNR) of the $m$th channel radar data cube in $\text{\eqref{eqn:measurement_hypotheses}}$. In our problem setting, SNR at the $k$th CPI for the $m$th channel is found as a function of the range bin $r$ and the object state $X_k$, i.e.,}
\begin{eqnarray}
&&{\mathrm{SNR}_{m,k}(r,X_{k})} \notag \\ 
&&{\,\,\,\,\triangleq \frac{\mathrm{E}\{\left(\alpha_{m,k}\mathbf{s}_{m}(r,X_k)\right)^{T}\left(\alpha_{m,k}\mathbf{s}_{m}(r,X_k)\right)\}}{\mathrm{E}\{\mathbf{n}_{m}(r)^{T}\mathbf{n}_{m}(r)\}}} \notag \\
&&{\,\,\,\,= \frac{\mathrm{E}\{\alpha_{m,k}^{*}\alpha_{m,k}\}\mathrm{E}\{\mathbf{s}_{m}(r,X_k)^{T}\mathbf{s}_{m}(r,X_k)\}}{\mathrm{tr}\{\Sigma_{m}\}},}
\label{eqn:m_th_SNR}
\end{eqnarray} 
{where $\alpha_{m,k} \triangleq \mathrm{Re}\{\alpha_{m,k}\} + j\mathrm{Im}\{\alpha_{m,k}\}$ is the complex reflection coefficient of the $m$th channel which is comprised of a real part (i.e., $\mathrm{Re}\{.\}$) and an imaginary part (i.e., $\mathrm{Im}\{.\}$), $\mathbf{s}_{m} \in \mathbb{C}^{LN \times 1}$ is the signal model associated with the range bin $r$ and the object state $X_k$ as given in $\text{\eqref{eqn:measurement_hypotheses}}$, and, $\mathrm{tr}\{\Sigma_{m}\}$ denotes the trance of $\Sigma_{m}$. Here, $\mathbf{n}_{m} \sim \mathcal{CN}(.; \boldsymbol{0}, \Sigma_{m})$ models the noise background of the $m$th channel and is a complex random variable with zero mean and covariance of $\Sigma_{m}$ as discussed in $\text{Section~\ref{sec:signal_model}}$.}

{We consider the SNR associated with the object state $X_k$ over the range bins in which, owing to the auto-correlation output $\Lambda_{m}$ in $\text{\eqref{eqn:column_data_model}}$, the second term in the nominator of $\text{\eqref{eqn:m_th_SNR}}$ yields}
\begin{eqnarray}
&&{\mathrm{E}\Big\lbrace\sum_{r \in {\cal E}_m({X_k})}\mathbf{s}_{m}(r,X_k)^{T}\mathbf{s}_{m}(r,X_k)\Big\rbrace} \notag\\
&&{\qquad\qquad\qquad = LN\times {\Lambda}_{m}({\cal E}_m({X_k}))} \\
&&{{\Lambda}_{m}({\cal E}_m({X_k})) \triangleq \sum_{r \in {\cal E}_m({X_k})} \Lambda^{*}_{m}(rT_{p}-\tau_{m}(X_k) - \Delta t_{m}) \notag }\\
&&{\qquad\qquad\qquad \times \Lambda_{m}(rT_{p}-\tau_{m}(X_k) - \Delta t_{m})},
\label{eqn:SNR_r_bins}
\end{eqnarray}
{where $L$ indicates the number of array elements, and $N$ is the number of transmitted pulses in a CPI. Thus, the SNR for the radar data cube at the $k$th CPI for the $m$th channel through $\text{\eqref{eqn:m_th_SNR}--\eqref{eqn:SNR_r_bins}}$ is given by}
\begin{eqnarray}
&&{\mathrm{SNR}_{m,k} =  \frac{ LN{\Lambda}_{m}({\cal E}_m({X_k}))\mathrm{E}\{\alpha_{m,k}^{*}\alpha_{m,k}\}}{\mathrm{tr}\{\Sigma_{m}\}}}
\label{eqn:m_th_SNR_r_bins} \\
&&{\mathrm{SNR}^{m,k}_{dB} = 10\log_{10}\left(\mathrm{SNR}_{m,k}\right),}
\label{eqn:m_th_SNR_r_bins_dB}
\end{eqnarray}
{where $\mathrm{SNR}^{m,k}_{dB}$ denotes $\mathrm{SNR}_{m,k}$ in the decibel (dB) $\text{\cite[Chp.6]{Richards2005b}}$.} 

{As a result, SNR for an integrated value of all radar data cubes up to $k$ CPIs for $M$ channels is found by using the summation of all the $m$th channel SNR, i.e.,}
\begin{eqnarray}
{\mathrm{SNR}_{k} =  \sum_{k^{'}=1}^{k}\sum_{m=1}^{M} \mathrm{SNR}_{m,k^{'}}.}
\label{eqn:SNR_k_integ}
\end{eqnarray}       

{Now, we explicitly show that the expectation of the long time likelihood ratio for the detection test equals to that of the $\mathrm{SNR}$ in $\text{\eqref{eqn:SNR_k_integ}}$. The test value at the $k$th CPI for detection, i.e., $\log L_{k}$ in $\text{\eqref{eqn:integration}}$, is found by using the summation of instantaneous likelihood ratios up to time $k$. The instantaneous likelihood ratio in $\text{\eqref{eqn:InslogL}}$ at time $k$ is easily factorised to the $m$th channel instantaneous likelihood ratio, i.e.,}  
\begin{eqnarray}
&&\mspace{-20mu}{L_{m}(\mathbf{Z}_{k}(X_k)|X_{k}, {\boldsymbol{\alpha}}_{k}, \Delta {\mathbf t}) \notag }\\
&&{\triangleq \sum_{r \in {\cal E}_m(X_k)} L(\mathbf{Z}_{m,k}(r)| X_{k}, \alpha_{m,k}, \Delta t_m)\notag} \\
&&{= \sum_{r \in {\cal E}_m({X}_{k})} \Big(2 \mathrm{Re}\big\{ {\alpha}^*_{m,k}\mathbf{s}_{m}(r,{X}_{k})^H \Sigma^{-1}_m \mathbf{Z}_{m,k}(r)\big\} \notag }\\
&&{\qquad -|{\alpha}_{m,k}|^2\mathbf{s}_{m}(r,{X}_{k})^H \Sigma^{-1}_m \mathbf{s}_{m}(r,{X}_{k})\Big).}
\label{eqn:InslgL}
\end{eqnarray}

{We take the expectation of this likelihood ratio and have} 
\begin{eqnarray}
&& \mspace{-20mu}{ \mathrm{E}\big\{L_{m}(\mathbf{Z}_{k}(X_k)|X_{k}, {\boldsymbol{\alpha}}_{k}, \Delta {\mathbf t})\big\} \notag }\\
&&{= \sum_{r \in {\cal E}_m({X}_{k})} \Big(2 \mathrm{Re}\big\{ {\alpha}^*_{m,k}\mathbf{s}_{m}(r,{X}_{k})^H \Sigma^{-1}_m \mathrm{E}\big\{\mathbf{Z}_{m,k}(r)\big\}\big\} \notag}\\
&&{\qquad-|{\alpha}_{m,k}|^2\mathbf{s}_{m}(r,{X}_{k})^H \Sigma^{-1}_m \mathbf{s}_{m}(r,{X}_{k})\Big).}
\label{eqn:Expectation_log_likelihood}
\end{eqnarray}

{From the radar data cube in $\text{\eqref{eqn:measurement_hypotheses}}$, when $H=H_1$ hypothesis holds, the expectation of $\mathbf{Z}_{m,k}(r)$ is given by}
\begin{eqnarray}
{\mathrm{E}\big\{\mathbf{Z}_{m,k}(r)\big\} = {\alpha}_{m,k}\mathbf{s}_{m}(r,{X}_{k}).} 
\label{eqn:Expectation}
\end{eqnarray}

{After substituting $\text{\eqref{eqn:Expectation}}$ into $\text{\eqref{eqn:Expectation_log_likelihood}}$, the resulting expression is found as}
\begin{eqnarray}
&& \mspace{-20mu}{\mathrm{E}\big\{L_{m}(\mathbf{Z}_{k}(X_k)|X_{k}, {\boldsymbol{\alpha}}_{k}, \Delta {\mathbf t})\big\} \notag }\\
&&{= \sum_{r \in {\cal E}_m({X}_{k})} |{\alpha}_{m,k}|^2\mathbf{s}_{m}(r,{X}_{k})^H \Sigma^{-1}_m \mathbf{s}_{m}(r,{X}_{k})} \label{eqn:SNR_} \\
&&{= \mathrm{SNR}_{m,k}(r,X_{k})}.
\end{eqnarray}
{As a result, the expectation of the $m$th instantaneous likelihood in $\text{\eqref{eqn:SNR_}}$ is equivalent to $\mathrm{SNR}_{m,k}(r,X_{k})$ in $\text{\eqref{eqn:m_th_SNR}}$. Therefore, the integrated value of $\log L_{k}$ in $\text{\eqref{eqn:integration}}$ is equivalent to an estimate of $\mathrm{SNR}_{k}$ in $\text{\eqref{eqn:SNR_k_integ}}$.}    

\section{Example}  
\label{sec:Example} 

\begin{table}[bt]
	\vspace{-1em}
	\renewcommand{\arraystretch}{1.3}
	\caption{Transmitted signal parameters}
	\centering
	\begin{tabular}[10pt]{c||c}
		\hline
		\bfseries Parameter & \bfseries Value\\
		\hline\hline
		Carrier frequency, i.e., $f_{c}$ & \SI{10}{\giga\hertz}\\
		Probing waveform bandwidth, i.e., $B$ & \SI{1}{\mega\hertz} \\
		Probing waveform duration, i.e., $T_p$ & \SI{1.0}{\micro\second}\\
		Pulse repetition interval (PRI), i.e., $T$ & \SI{100}{\micro\second} \\
		Number of range bins, i.e., $\Gamma$ & 100\\
		Number of pulses, i.e., $N$ & $20$\\
		Number of elements in the ULA, i.e., $L$ & $20$ \\
		Length of the coherent processing interval (CPI) & \SI{2}{\milli\second}\\
		Illumination period ($\Delta$ in~\eqref{eqn:Markov_transition})   & \SI{0.1}{\second}\\
		Number of transmitters, i.e., $M$ & $2$ \\	
		\hline
	\end{tabular}
	\label{tab:transmitted_pulse_parameter}
\end{table}

\begin{figure}[bt]
	\subfigure[{Problem scenario}]{\includegraphics[width=1.7in]{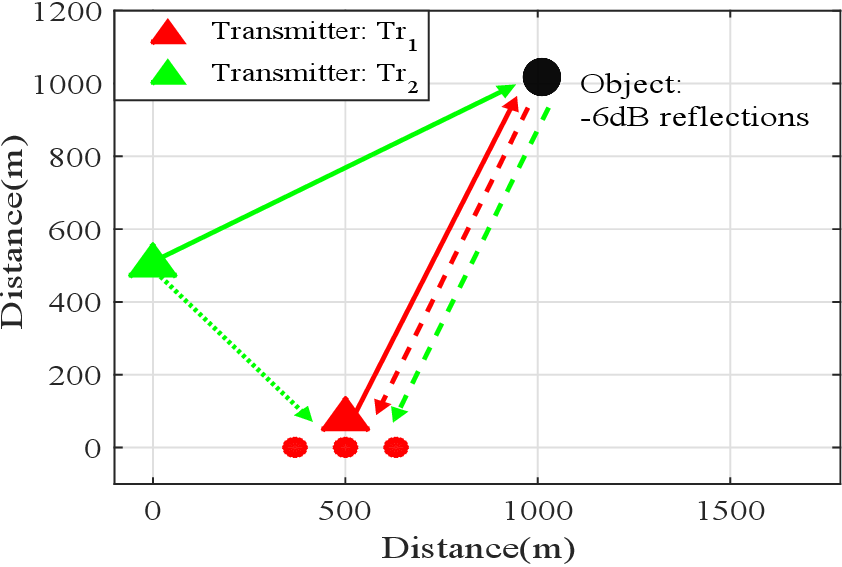}}
	\subfigure[{Object's trajectory}]{\includegraphics[width=1.7in]{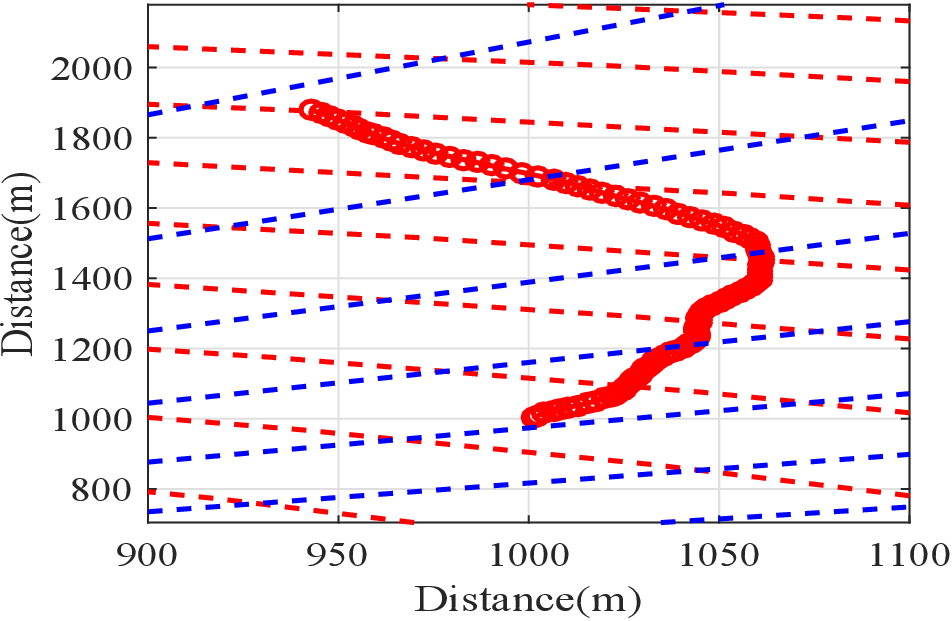}}
	\centering
	\vspace{-0.5em}
	\caption{{Example scenario: (a) $M=2$ transmitters (i.e., $\mathrm{Tr}_{1}$ and $\mathrm{Tr}_{2}$) emitting $N=20$ pulses (solid line arrows) towards an small object (a black dot). A ULA (red dots) collects low SNR (-6dB) reflections (dashed line arrows) and direct signals (a dotted green line arrow). (b) The object's trajectory depicted with the red line. The range bins resulting from sampling in time is shown by the dashed red lines. The bearing bins of the conventional processing chains is shown by the dashed blue lines.}}
	\vspace{-1.2em}
	\label{fig:example_of_scenario}
\end{figure}

In this section, we demonstrate the proposed algorithm through an example and compare the efficacy of this approach with conventional techniques. {We consider a scenario in which a ULA (red dots) receiver co-located with a transmitter (red triangle) is at $[500\mathrm{m},0\mathrm{m}]$ of the 2D Cartesian plane, and, a separated transmitter (green triangle) is located at $[0\mathrm{m},500\mathrm{m}]$ (see. $\text{\figurename~\ref{fig:example_of_scenario}}$). In this setting, $M=2$ transmitters emit $N=20$ linear frequency modulated (i.e., up-chirp) waveforms towards a surveillance region and repeats this illumination pattern every $\SI{0.1}{\second}$. In this region, there is a small object (black dot) with an initial state $X_0 = [1000\mathrm{m},1000\mathrm{m}, 10\mathrm{m/s}, 50\mathrm{m/s}]$ moving along an unknown trajectory (red line) generated from the object dynamic model defined in $\text{\eqref{eqn:Markov_transition}}$.} The ULA receiver with $L=20$ elements collects measurements (dashed line arrows) in accordance with the signal model in~\eqref{eqn:measurement_hypotheses} from the local (dashed red line arrow) and the remote (dashed green line arrow) channels. Superpositioned in the remote channel is the direct probing transmission from the transmitter. The parameter configuration of these transmissions are shown in Table~\ref{tab:transmitted_pulse_parameter}. 

We simulate $100$ independent sets of trajectories, and array measurements: {When the $H=H_{1}$ hypothesis holds, the array measurements at the $k$th CPI are associated with the object state $X_k$ and the reflection coefficient $\alpha_{m,k}$. These quantities are generated from a complex Gaussian by using}
\begin{eqnarray}
	{\mathbf{Z}_{m,k}(r)}&&{\sim \mathcal{CN}(.;\alpha_{m}\mathbf{s}_{m}(r,X_{k}),\Sigma_{m}), }\\	
	&&{m = 1,\dots, M,\,\ r \in {\cal E}(X_{k}),}\notag
\end{eqnarray} 
{where $m$ indicates the $m$th channel, ${\cal E}(X_{k})$ is a set of the range bins associated with $X_k$ in $\text{\eqref{eqn:setOfrange}}$}.

{Otherwise, the measurements are generated from} 
\begin{eqnarray}
{\mathbf{Z}_{m,k}(r)} &&{\sim \mathcal{CN}(.;\mathbf{0},\Sigma_{m}),}\\ 
&&{m = 1,\dots, M,\,\ r \in \Gamma\backslash{\cal E}(X_{k}),} \notag
\end{eqnarray}
{where $\Gamma$ is the length of range bins in $\text{\eqref{eqn:mthsampled}}$.  
Here, the expected SNR for the $m$th channel measurement in a CPI is $-6\mathrm{dB}$. This quantity is found by using $\mathrm{SNR}^{m,k}_{dB}$ in $\text{\eqref{eqn:m_th_SNR_r_bins_dB}}$.}        

The direct signal from the non co-located transmitter is received with additive noise using \eqref{eqn:direct channel} with an SNR of $0\mathrm{dB}$. The time reference shift of the remote transmitter and the receiver, i.e., $\Delta t$, is selected randomly in the range of $0 < \Delta t < \mathrm{PRI}$, and, this value is used for all experiments.

{We use $\text{Algorithm~\ref{Algorithm:proposed_long_time_algorithm}}$ for $100$ CPIs that spans $\SI{10}{\second}$ which contains $100$ CPIs. Each CPI corresponds to one radar data cube (see, $\text{\figurename~\ref{fig:Data_acquisition}}$). We compare the performance of our algorithm in this scenario with the following detectors:}
\begin{enumerate}
 \item The clairvoyant detector: This detector uses the ground truth values of the unknown parameters (i.e., the object trajectories, reflectivities, and, the synchronisation term) when evaluating the logarithm of the likelihood ratio test in~\eqref{eqn:LRT}. {In other words, this test substitutes the true values of the unknowns in $\text{\eqref{eqn:integration}}$ and leads to }
 \begin{eqnarray}
 &&{\log L_k \underset{H_{0}}{\overset{H_{1}}{\gtrless}} \log \mathcal{T}_{k}}  \\
 &&{\log L_k = \log L_{k-1} \notag}\\
 &&{\quad + L(\mathbf{Z}_{k}(X_{\text{true},k})|X_{\text{true},k}, {\boldsymbol{\alpha}}_{\text{true},k}, \Delta {\mathbf t}_{\text{true}})},
 \end{eqnarray} 
{where $X_{\text{true},k}$, ${\boldsymbol{\alpha}}_{\text{true},k}$, and $\Delta {\mathbf t}_{\text{true}}$ are the true values of $X_{k}$, $\boldsymbol{\alpha}_{k}$, and $\Delta \mathbf{t}$, respectively.} 

 The CFAR threshold, i.e, $\log \mathcal{T}_{k}$, for this detector is found using \eqref{eqn:knownsignal}--\eqref{eqn:CFAR_Threshold} as discussed in Section~\ref{sec:CFARThreshold}.
 
 {Note that the clairvoyant detector is the optimal detector~$\text{\cite[Chp.13]{Kay1998}}$. The $k$ integrated value of $\log L_k$ with the ground truth values provides the maximum achievable value for the detection test. Therefore, we use this integrated value as the performance upper bound when comparing the efficacy of the proposed integration approach in this section.}  

 \item Conventional coherent detector: This detector processes the measurements after mapping them over a grid of bearing and Doppler bins. These bins correspond to resolution cells which are found for the example system as follows: The bearing resolution is found as $\Delta \theta = 5.1^{\circ}$ using $\Delta \theta = \sin^{-1}\big(\frac{0.8192}{L}\big)$. The range resolution is found as $\Delta\tau = 150\text{m}$ using $\Delta\tau=\frac{c}{2B}$ (see, e.g., \cite{VanTrees2002m}). The velocity resolution of the conventional processing is found as $\Delta{V} = 7.5\mathrm{m/s}$ by using $\Delta{V}=\frac{\lambda_c}{2NT}$ (or, equivalently, the Doppler resolution $\Delta \omega = 4\pi f_c\frac{\Delta{V}}{c}T$ as $\pi/10$~\SI{}{\radian\per\second}). This detector integrates the mapped complex values for the same ``cell under test'' across time without taking account object manoeuvres~\cite{Richards2005b}.
\end{enumerate}

We initiate our algorithm with $P=400$ particles as a $20 \times 20$ uniform grid over a bounded region of location and velocity vectors such that the locations span the {``cell under test''}~\footnote{Note that because all the steering vectors during processing are selected by the Bayesian recursive filtering there is no fixed bearing or velocity resolution cells for our approach unlike the conventional detector.}. As the Bayesian filtering and trajectory estimation steps iterate, these particles evolve to converge to the true state of the object simultaneously giving rise to the integrated value in \eqref{eqn:integration}.

\begin{figure}[bt]
	\includegraphics[width=3in]{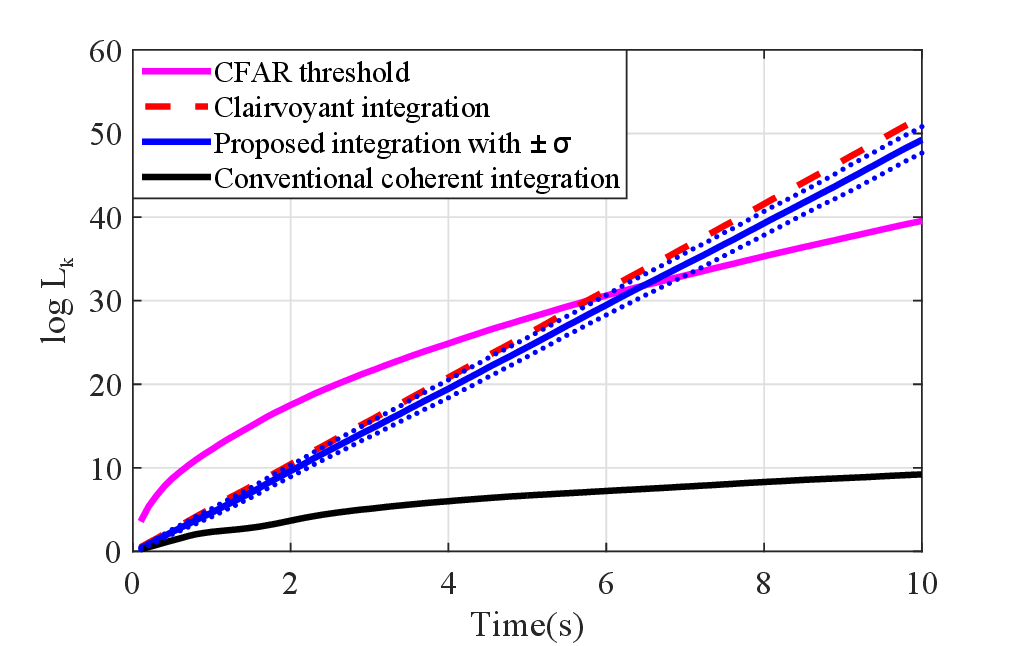}
	\centering
	\vspace{-0.5em}
	\caption{Long-time integration using the proposed scheme, the clairvoyant integrator, and the conventional coherent integrator: The integrated sufficient statistics from the proposed integration averaged over $100$ experiments is depicted by the solid blue line. The integrated value from the clairvoyant integrator is the dashed red line and {the clairvoyant (CFAR) threshold for $P_{fa} = 10^{-6}$ (averaged for $100$ experiments)} is the solid magenta line. The conventional scheme leads to the solid black line.}
	\vspace{-1.2em}
	\label{fig:Result_of_direct_channel_1}
\end{figure}

\begin{figure}[bt]
	\subfigure[Long-time integration with $M=2$ transmitters]{\includegraphics[width=3in]{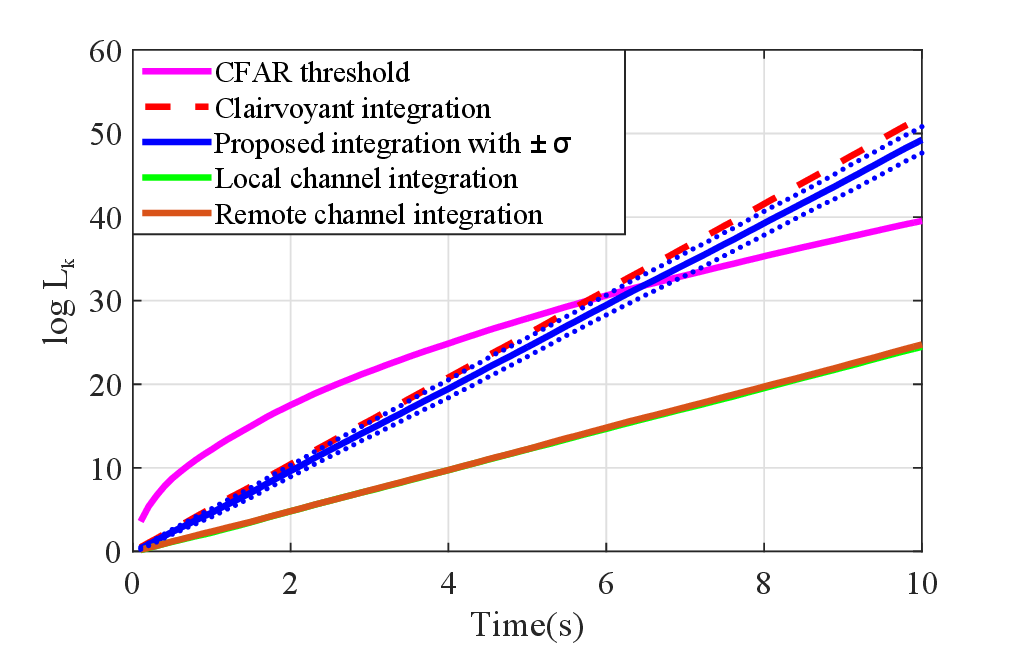}}
	\subfigure[{Long-time integration with $M=4$ transmitters}]{\includegraphics[width=3in]{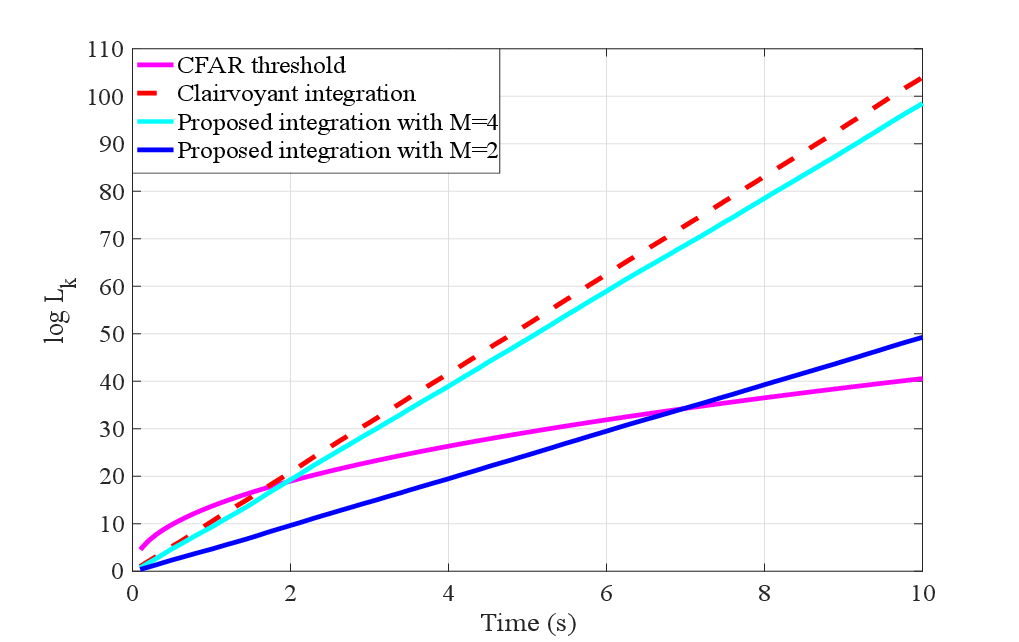}}
	\centering
	\vspace{-0.5em}
	\caption{The proposed scheme (solid blue line) versus the single channel integrations: (a) The proposed scheme (solid blue line) with $M=2$ transmitters. The local channel (solid green line) integration, and the remote channel (solid brown line) integration fail to exceed the detection threshold. (b) The proposed scheme (solid cyan line) with $M=4$ transmitters. {The clairvoyant (CFAR) threshold for $P_{fa} = 10^{-6}$ (averaged for $100$ experiments) is the solid magenta line in both (a) and (b).}}
	\vspace{-1.2em}
	\label{fig:Result_of_direct_channel_2}
\end{figure}

\subsection{{Detection test via long time integration}}
{We first consider the proposed long time integration for the detection test as discussed in $\text{Section~\ref{sec:Long_time_integration}}$. We repeatedly use $\text{Algorithm~\ref{Algorithm:proposed_long_time_algorithm}}$ with $100$ scenario realisations, and, compare the resulting long time integration performance with that of the clairvoyant and the conventional detector. In~$\text{\figurename~\ref{fig:Result_of_direct_channel_1}}$, the integration values are given as a function of time. The clairvoyant integrator sets an upper bound for the integrated sufficient statistics, the average of which is depicted by the dashed red line. Long time integration accuracy of the proposed algorithm is coupled to the trajectory estimation performance through the EM iterations for finding the reflection coefficients (i.e., $\text{Algorithm~\ref{Algorithm:EM_algorithm}}$). In $\text{\figurename~\ref{fig:Result_of_direct_channel_1}}$, the proposed scheme's performance is very close to the clairvoyant detector bound (solid blue line rendering the average with $\pm \sigma$ bounds shown with dotted blue lines). Here, the proposed integration reaches to $49.24$ at $t=10\mathrm{s}$, which is relatively close to $51.78$ achieved by the clairvoyant integration. This indicates that the loss in integration performance due to estimation errors of the target trajectory and reflection coefficients is very small. The conventional scheme fails to continue the integration after the object leaves the initial cell under test. This integration is shown with the solid black line in $\text{\figurename~\ref{fig:Result_of_direct_channel_1}}$.} 

{The clairvoyant CFAR detection threshold for $P_{fa}=10^{-6}$ is depicted as the solid magenta line (averaged for the $100$ experiments) in $\text{\figurename~\ref{fig:Result_of_direct_channel_1}}$. Detection for each detector is made by comparing its integration value against this threshold. The proposed scheme exceeds the CFAR threshold and enables us to decide on the object existence hypothesis ($H=H_{1}$) at $t = 6.5\mathrm{s}$ whereas the conventional scheme stays in the region for the noise only signal hypothesis ($H=H_{0}$).}     

{$\text{\figurename~\ref{fig:Result_of_direct_channel_2}(a)}$ compares the proposed multi-channel integration with the integration using measurements in only one of the $M=2$ channels. In other words, $\text{Algorithm~\ref{Algorithm:proposed_long_time_algorithm}}$ is used only with the data cube from the local channel (solid green line) and that from the remote channel (solid brown line), respectively. The results are averaged over $100$ experiments. All integrations (i.e., the multi-channel integration, the local channel integration, and the remote channel integration) increase over time, however, both the local channel and the remote channel integration fail to exceed the CFAR threshold by themselves. $\text{\figurename~\ref{fig:Result_of_direct_channel_2}(b)}$ illustrates the proposed integration with $M=4$ transmitters (solid cyan line) in comparison with the previous integration for $M=2$ transmitters (solid blue line). $\text{Algorithm~\ref{Algorithm:proposed_long_time_algorithm}}$ with $M=4$ transmitters exceeds the CFAR threshold at $t = 2\mathrm{s}$ which is less than half of the time required for $M=2$ transmitter case ($t=6.5\mathrm{s}$) revealing the advantage of using more transmitters.}  

\begin{figure}[bt]
	\subfigure[$P_d$ for the proposed scheme with $M=2$ transmitters]{\includegraphics[width=3in]{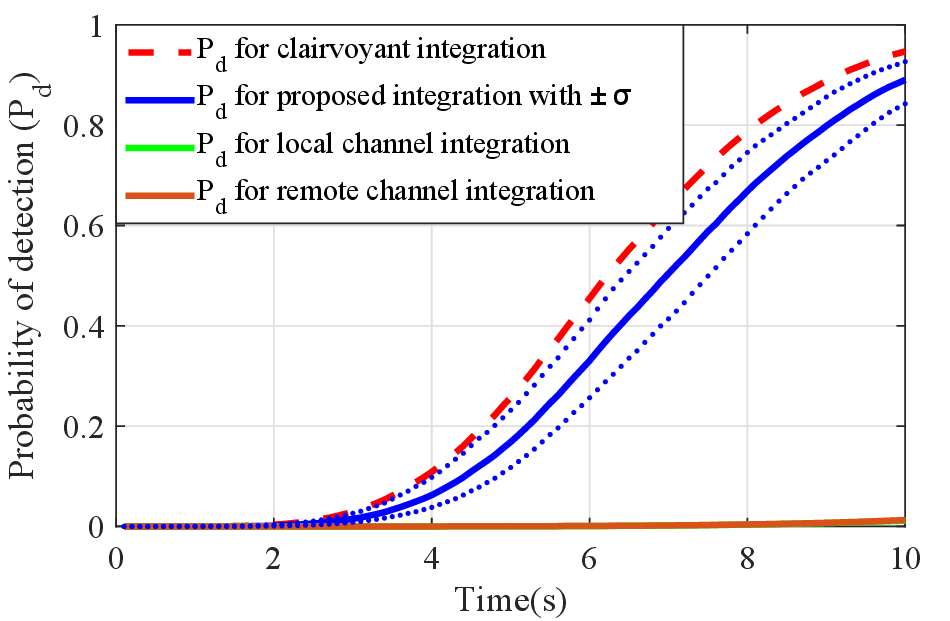}}
	\subfigure[{$P_d$ for the proposed scheme with $M=4$ transmitters}]{\includegraphics[width=3in]{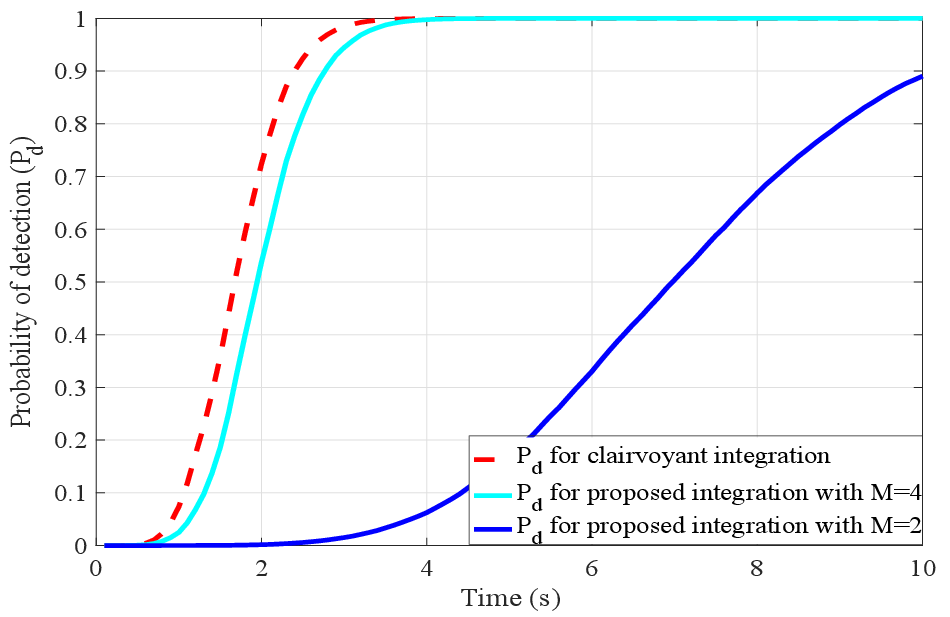}}
	\centering
	\vspace{-0.5em}
	\caption{Probability of detection $(P_d)$ for the proposed scheme in comparison with the clairvoyant detector and the conventional detector: (a) $P_d$ for the proposed scheme (solid blue line) with $M=2$ transmitters. (b) $P_d$ for the proposed scheme (solid cyan line) with $M=4$ transmitters compared to $P_d$ for the clairvoyant detector (dashed red line). {The probability of false alarm $P_{fa} = 10^{-6}$ compared to $P_d$ for the clairvoyant detector (dashed red line) in both (a) and (b).}}
	\vspace{-1.2em}
	\label{fig:PD}
\end{figure}

\begin{figure}[bt]
	{\includegraphics[width=3in]{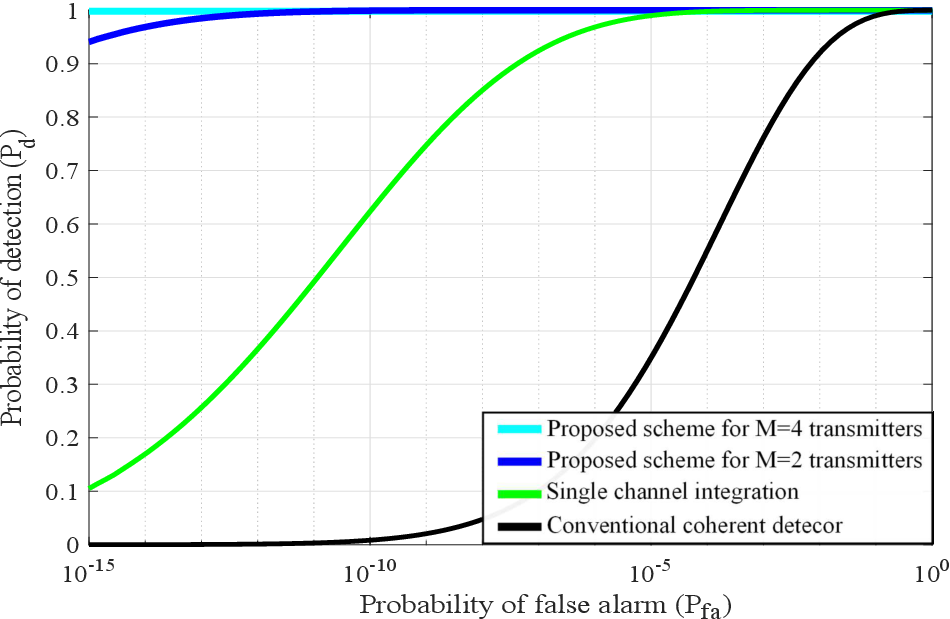}}
	\centering
	\vspace{-0.5em}
	\caption{{Receiver operating characteristic (ROC) curves: The proposed detector with $M=4$ transmitters (solid cyan line), and, $M=2$ transmitters (solid blue line), respectively, are given. The green and the black lines denote the single channel and the conventional coherent integrations, respectively.}}
	\vspace{-1.2em}
	\label{fig:ROC}
\end{figure}

{Next, we consider the probability of detection $P_d$ as a function of the integration time in~$\text{\figurename~\ref{fig:PD}}$. Here, the $P_d$ of the proposed scheme is found empirically and is averaged over the experiments. The $P_d$ of the clairvoyant detector (dashed red line) sets the upper performance bound. The $P_d$ of the proposed scheme for $M=2$ transmitters is drawn by the solid blue line in~$\text{\figurename~\ref{fig:PD}(a)}$. This quantity increases with time and reaches $P_d=0.89$ at $t=10\mathrm{s}$, which is relatively close to the $P_d=0.95$ of the clairvoyant detector. The $P_{d}$ functions of the local channel (solid green line) and the remote channel integration (solid brown line) stay in the vicinity of zeros, and, indicate that they fail to detect the object in an overwhelming majority of the experiments. $\text{\figurename~\ref{fig:PD}(b)}$ illustrates the $P_d$ of the proposed scheme for $M=4$ transmitters (solid cyan line). This quantity increases with time and reaches $P_d=1$ at $t=4\mathrm{s}$, which enables us to detect the object much faster than the value used in~$\text{\figurename~\ref{fig:PD}(a)}$.}

{Now, we consider the probability of detection $P_d$ as a function of different false alarm ($P_{fa}$) values in the range of $P_{fa} = 10^{0}$ and $P_{fa} = 10^{-15}$. This is illustrated in~$\text{\figurename~\ref{fig:ROC}}$. and referred to as receiver operating characteristic (ROC) curve~$\text{\cite[Chp.3]{Kay1998}}$. We fix the integration time to $t=10\mathrm{s}$ for ROC calculation as $P_{d}$ is also a function of integration time (see $\text{\figurename~\ref{fig:Result_of_direct_channel_2}}$). We compare the ROCs obtained by using $\text{Algorithm~\ref{Algorithm:proposed_long_time_algorithm}}$ for $M=4$ and $M=2$ channels, respectively, and a single channel with the ROC of the conventional coherent detector (solid black line). The ROC of the proposed integration for $M=4$ transmitters (solid cyan line) provides $P_d = 1$ after $P_{fa} = 10^{-15}$, whereas the integration value for $M=2$ transmitters (solid blue line) yields $P_d = 1$ after $P_{fa} = 10^{-13}$. Furthermore, the single channel integration (solid green line) enables us to have $P_d = 1$ above a small false alarm rate of $P_{fa} = 10^{-5}$. The conventional coherent integration, however, provides $P_d = 1$ after $P_{fa} = 10^{-1}$.}           

%

\subsection{{Performance in estimating the unknowns}}
\begin{figure}[t]
	\subfigure[Typical scenario for an estimated trajectory]{\includegraphics[width=1.7in]{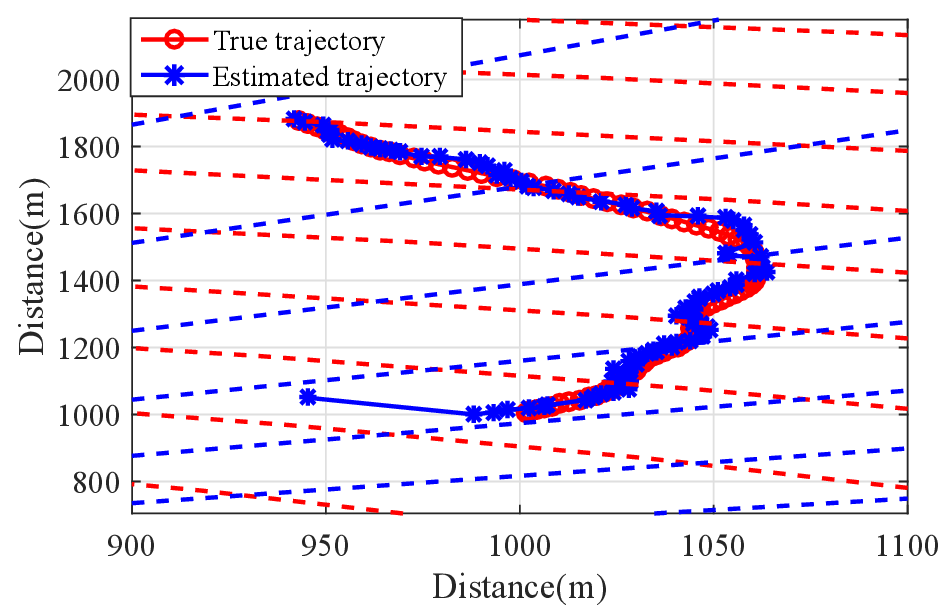}}
	\subfigure[{RMSE of the range estimation}]{\includegraphics[width=1.7in]{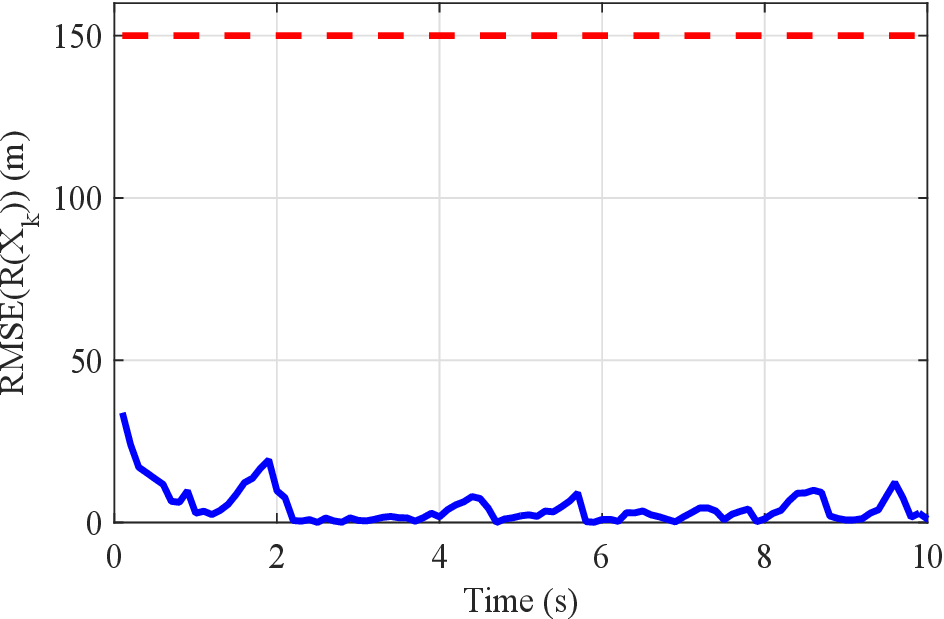}}
	\subfigure[{RMSE of the velocity estimation}]{\includegraphics[width=1.7in]{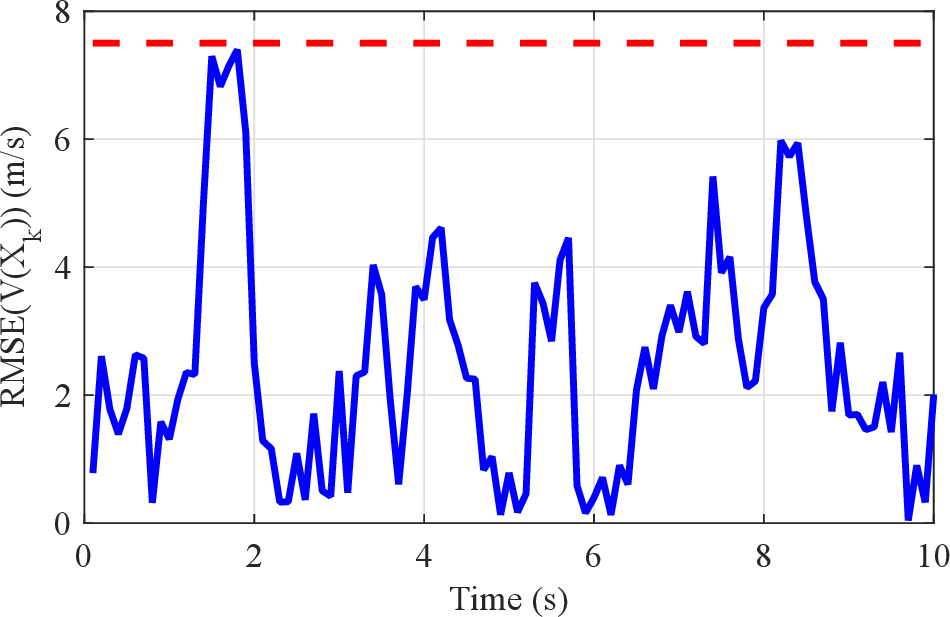}}
	\subfigure[{RMSE of the bearing estimation}]{\includegraphics[width=1.7in]{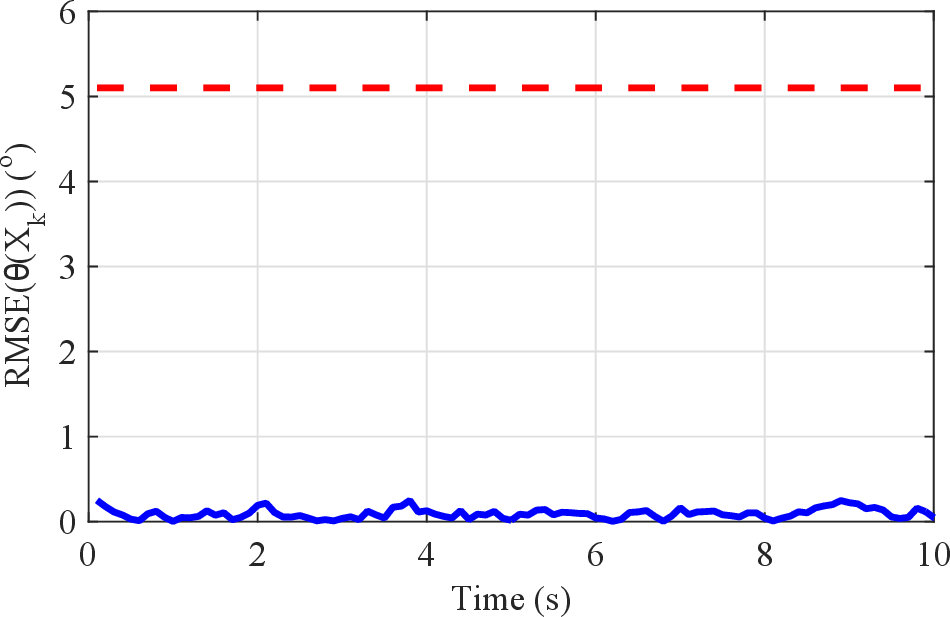}}
	\centering
	\vspace{-0.5em}
	\caption{{Typical trajectory estimation: (a) The estimated trajectory by the proposed algorithm is depicted with the blue line. (b) Root mean square error (RMSE) of the range estimation in (a). (c) RMSE of the velocity estimation in (a). (d) RMSE of the angle of arrival estimation in (a). The dashed red lines in (b), (c), and (d) are the range resolution ($\Delta \tau = 150 \mathrm{m}$), the velocity resolution ($\Delta V = 7.5 \mathrm{m/s}$) and the bearing resolution ($\Delta \theta = 5.1^{\circ}$) }}
	\vspace{-1.2em}
	\label{fig:Estiamteion}
\end{figure}

{Here, we demonstrate the inner workings of $\text{Algorithm~\ref{Algorithm:proposed_long_time_algorithm}}$. In particular, we consider the estimation accuracy of the object trajectory, the reflectivities, and, the synchronisation term within $\text{Algorithm~\ref{Algorithm:proposed_long_time_algorithm}}$. $\text{\figurename~\ref{fig:Estiamteion}(a)}$ illustrates a typical trajectory (red line) which would lead to resolution bin migrations in conventional processing. The trajectory estimate output by the proposed algorithm is depicted as the blue line along with the resolution bins (dashed lines) of a conventional detector. 
$\text{\figurename~\ref{fig:Estiamteion}(b)}$ shows the root mean squared error (RMSE) of the corresponding range estimate in comparison with the range resolution of $\Delta \tau$ (dashed red line). Note that the error reduces to the $3.3 \%$ of the range resolution after $\SI{2.3}{\second}$. $\text{\figurename~\ref{fig:Estiamteion}(c)}$ presents the RMSE of the velocity component of the trajectory estimate in~$\text{\figurename~\ref{fig:Estiamteion}(a)}$. This estimate error is below the velocity resolution bin of $\Delta{V}$ (dashed red line), where the error between $\SI{1}{\second}$ and $\SI{2}{\second}$ shows a relatively large value due to the object's manoeuvres. $\text{\figurename~\ref{fig:Estiamteion}(d)}$ illustrates the RMSE of the bearing component of the trajectory estimate in $\text{\figurename~\ref{fig:Estiamteion}(a)}$. Here, the estimate error is a very small value compared to the bearing resolution of $\Delta\theta$ (dashed red line). Note that the resolution bins of the system provides only a coarse view of the trajectory whereas the proposed algorithm yields a super-resolution effect.}

\begin{figure}[t]
	\subfigure[{Reflection coefficient for local channel}]{\includegraphics[width=1.7in]{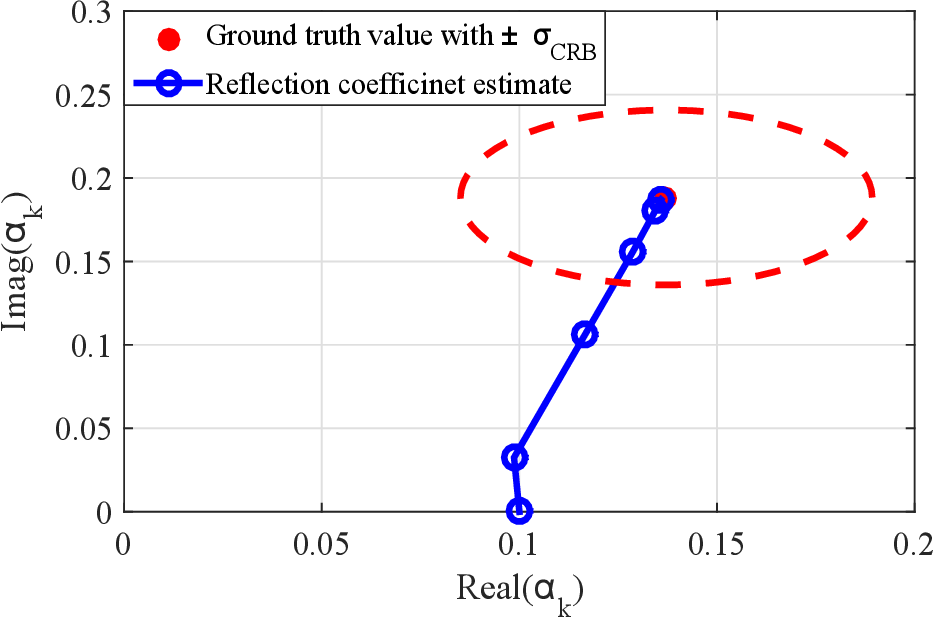}}
	\subfigure[{Reflection coefficient for remote channel}]{\includegraphics[width=1.7in]{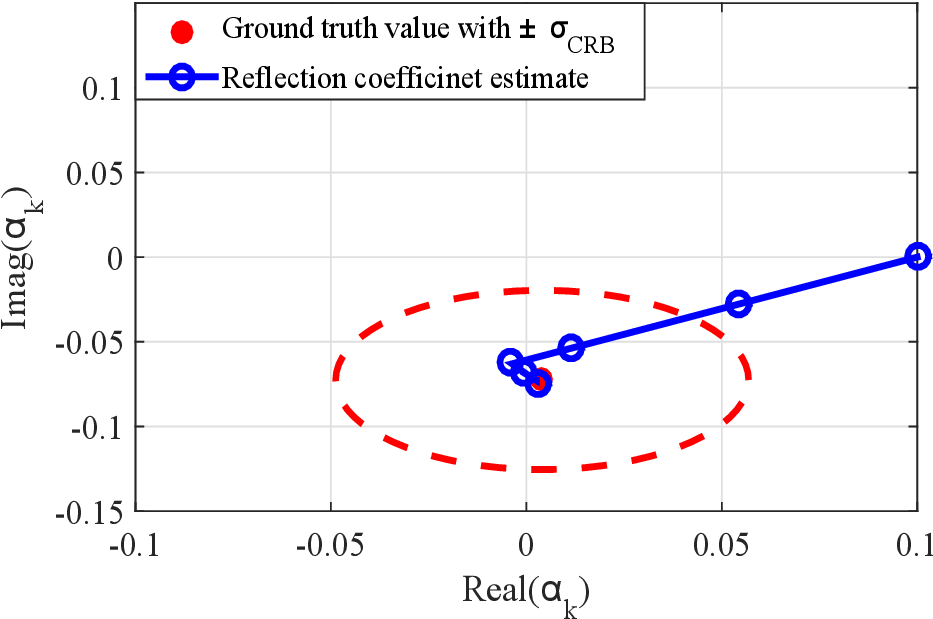}}
	\centering
	\vspace{-0.5em}
	\caption{{Complex reflection coefficient estimation with $-6\text{dB}$ reflections: (a) A typical estimate of the complex reflection coefficient for the local channel by using the proposed algorithm. The blue line indicates typical estimates of the local reflection coefficient by using $\text{Algorithm~\ref{Algorithm:EM_algorithm}}$ within $\text{Algorithm~\ref{Algorithm:proposed_long_time_algorithm}}$. The blue circles show $i=6$ iterations for finding it. The resulting estimate is compared to the ground truth value (red dot) with the $\pm$ standard deviation of Cram\'{e}r-Rao bound (CRB), i.e., $\pm \sigma_{\text{CRB}}$ (dashed red ellipse). The $x$ axis denotes the real part of the complex reflection coefficient and the $y$ axis is its imaginary part. (b) A typical estimate of the complex reflection coefficient for the remote channel by using $\text{Algorithm~\ref{Algorithm:EM_algorithm}}$ within $\text{Algorithm~\ref{Algorithm:proposed_long_time_algorithm}}$ with the same colour codes in (a).}}
	\vspace{-1.2em} 
	\label{fig:reflection_estimation}
\end{figure}

\begin{figure}[t]
	\centering
	\includegraphics[width=3in]{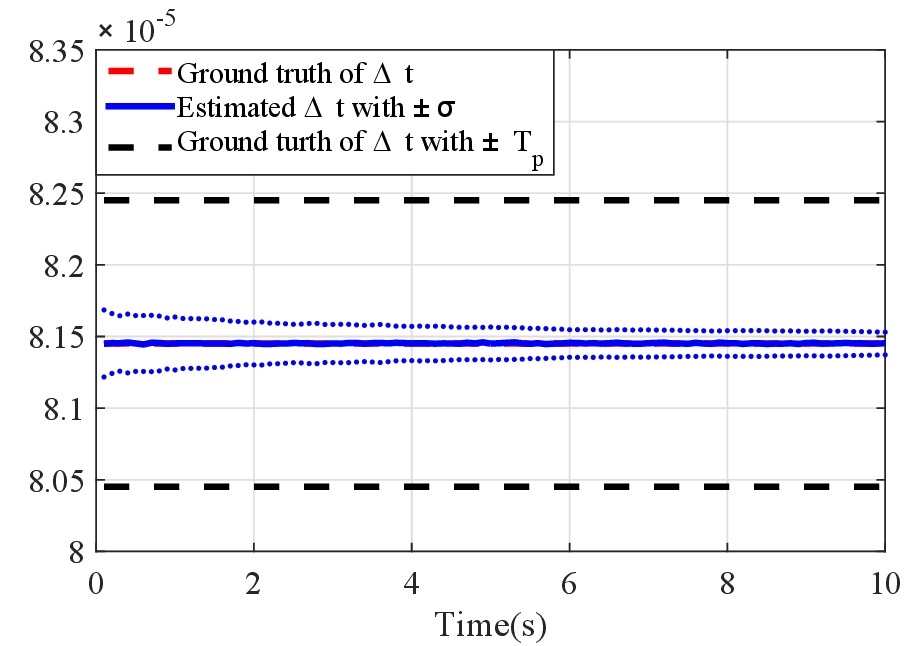}	
	\centering
	\vspace{-0.5em}
	\caption{Synchronised term estimation: Averaged synchronised term (solid blue line) estimated by using the proposed estimator versus the ground truth value (solid red line) with the bound the $\pm T_p$ bound (dashed black lines) of a preliminary search over the grids.}
	\vspace{-1.2em}
	\label{fig:RMSE_Synchro}
\end{figure}

{Next, we consider the estimation performance in finding the complex reflection coefficients in the radar data cube. For this purpose, we use~$\text{Algorithm~\ref{Algorithm:EM_algorithm}}$ within~$\text{Algorithm~\ref{Algorithm:proposed_long_time_algorithm}}$. $\text{\figurename~\ref{fig:reflection_estimation}}$ shows typical estimates of the complex reflection coefficients for the typical steps of $\text{Algorithm~\ref{Algorithm:EM_algorithm}}$, where the $x$ axis indicates the real part of the complex reflection coefficient, and, the $y$ axis shows its imaginary part. We compare the resulting estimates with their ground truth values. Also given are the $\pm$ standard deviation of Cram\'{e}r-Rao bound (CRB), i.e., $\pm \sigma_{\text{CRB}}$ (see, the derivation of CRB in~$\text{Appendix~\ref{app:Jeffrey})}$ for comparison. In~$\text{\figurename~\ref{fig:reflection_estimation}(a)}$, the estimated reflection coefficient (blue line) for the local channel stays within $\pm \sigma_{\text{CRB}}$ (dashed red ellipse) after only a few iterations (solid blue circles), where the solid blue circles indicate the number of $i$ iterations for finding the reflection coefficient in~$\text{Algorithm~\ref{Algorithm:EM_algorithm}}$. The resulting estimate is close to its ground truth value (red dot). For the remote channel, $\text{\figurename~\ref{fig:reflection_estimation}(b)}$ presents a typical estimate of the remote complex reflection coefficient. The resulting estimate for the remote channel (solid blue line) stays within $\pm \sigma_{\text{CRB}}$ (dashed red ellipse), and, is close to the ground truth value (red dot).  
Note that both the local and the remote reflection coefficients estimated by the proposed algorithm are close to the ground truth values. It is also seen that these estimation errors stay within $\sigma_{\text{CRB}}$ after a few iterations.}

Now, we consider the estimation of the time shift~$\Delta t$ in the remote channel. For this purpose, we use~$\text{Algorithm~\ref{Algorithm:Golden_section_search_1}}$ within ~$\text{Algorithm~\ref{Algorithm:proposed_long_time_algorithm}}$ for the $100$ realisations. $\text{\figurename~\ref{fig:RMSE_Synchro}}$ presents the averaged estimates (solid blue line) with $\pm \sigma$ bounds (dotted blue lines). We compare these values with the true value of $\Delta t$ (red solid line). Also given are the  $\pm T_p$ bounds (dashed black lines) for comparison. It is seen that the estimation error stays within a small fraction of the total pulse width $T_p$.

{The benefits of our scheme come with some additive cost of computations compared to conventional scheme. The computational complexity of our algorithm for the cell under test for $K$ CPIs is $\mathcal{O}(PN_{r(X)}N_{I}M(LN)^2)$, whereas the conventional coherent detector requires $\mathcal{O}(KM(LN)^2)$. Here, $P$ indicates the number of particles, $LN$ is the length of the measurement vector in~$\text{\eqref{eqn:measurement_hypotheses}}$, $N_{r(X)}$ indicate length of ${\cal E}(X_{k})$, and $N_{I}$ denotes the number of iterations for the EM algorithm in~$\text{Algorithm~\ref{Algorithm:EM_algorithm}}$. 
}  

\section{Conclusion}
\label{sec:Conclusion}
In this work, we have proposed a detection algorithm that performs the most efficient statistical test in order to detect manoeuvring and low SNR objects with an arbitrarily long time window of measurements. This test is carried out by simultaneous trajectory estimation and long time integration. Our approach can operate in mono-static, bi-static, and multi-static configurations, and, enables us to collect the entire evidence of object existence at the receiver by i)~coherently integrating both mono-static and bi-static channels within a CPI, ii)~performing non-coherent integration across different channels, and, iii)~continuing integration for an arbitrarily long interval that contains many CPIs. 

We have demonstrated that our approach can provide integration for an arbitrarily long interval with an effectiveness close to the best achievable by a clairvoyant integrator. As a result, this approach enables us to detect manoeuvring and very low SNR objects which cannot be detected by using conventional techniques.


%
\appendices
\section{Likelihood locality}
\label{app:Likelihood locality}
Let us consider the likelihood ratio test in~\eqref{eqn:LRT_all} with the locality of the measurements ${\vect{Z}_{m,k}}(r) \in {\cal E}_m(X_k)$ to $X_k$. Let us define the complement of $\cal E$ in the set of range bins, i.e., $\bar {\cal E}\triangleq \{1,2,\cdots,\Gamma\} \setminus {\cal E}_m(X_k)$. It can easily be seen that 
\begin{eqnarray}
 &l&(\vect{Z}_{m,k}| X_k, \alpha_{m,k}, \Delta t_m, H=H_1)= \notag\\
 &&\prod_{r \in {\cal E}_m} l(\vect{Z}_{m,k}(r)| X_k, \alpha_{m,k}, \Delta t_m, H=H_1) \prod_{r' \in \bar {\cal E}_m} p( \vect{Z}_{m,k}(r') ). \notag\\
\label{eqn:h1fact}
\end{eqnarray}
Similarly, the likelihood for the noise-only signal hypothesis factorises as
\begin{eqnarray}
 &l&(\vect{Z}_{m,k}| X_k, H=H_0) = \notag\\
 &&\prod_{r \in {\cal E}_m} l(\vect{Z}_{m,k}(r)| H=H_0) \prod_{r' \in \bar {\cal E}_m} p( \vect{Z}_{m,k}(r') ), 
\end{eqnarray}
which, after substituting into~\eqref{eqn:LRT_all} with~\eqref{eqn:h1fact} leads to~\eqref{eqn:LRT}.

\section{{Cram\'{e}r-Rao bound (CRB) for complex reflection coefficients}}
\label{app:Jeffrey}
{Let us consider the Cram\'{e}r-Rao bound (CRB) for the complex reflection coefficients estimated by $\text{Algorithm~\ref{Algorithm:EM_algorithm}}$. The CRB provides the theoretical minimum variance for an unbiased estimator, and, is found by using inverse Fisher information~$\text{\cite[Chp.3]{Kay1993}}$. In our problem setting, the Fisher information is found by taking the second order partial derivative of the logarithm of the likelihood with respect to the reflection coefficient, i.e.,}
\begin{eqnarray}
{\mathrm{\mathbf{I}}(\alpha_{m,k})}&&{=
-\mathrm{E}\left\lbrace \frac{\partial^2 \log l(\mathbf{Z}_{k}|\boldsymbol{\alpha}_{k})}{\partial \alpha_{m,k}^2} \right\rbrace, \label{eqn:Fisher_Information}} \\ 
{\log l(\mathbf{Z}_{k}|\boldsymbol{\alpha}_{k})} &&{= \log \Big\lbrace \int_{X_{k}}\int_{\Delta\bf{t}}l(\mathbf{Z}_{k}|X_{k}, \boldsymbol{\alpha}_{k}, \Delta\mathbf{t}) \notag}\\ 
&&{\quad\times p(X_{k}, \Delta\mathbf{t}|\mathbf{Z}_{1:k-1})\mathrm{d}X_{k}\mathrm{d}\Delta\mathbf{t}\Big\rbrace,}
\label{eqn:log_likelihood_alpha} 
\end{eqnarray}
{where $\mathbf{I}(\mathbf{\alpha_{m,k}})$ denotes the Fisher information of the $m$th reflection coefficient at the $k$th CPI, and $\mathrm{E}\{\cdot\}$ is the expectation of its input argument.} 

{In order to evaluate $\log l(.)$ in~$\text{\eqref{eqn:log_likelihood_alpha}}$, we use the ground truth values of the object state $X_{k}$ and the synchronisation term $\Delta\mathbf{t}$. After substituting these true values into~$\text{\eqref{eqn:log_likelihood_alpha}}$, the resulting expression is found as}
\begin{eqnarray} 
&&{\log l(\mathbf{Z}_{k}|\boldsymbol{\alpha}_{k}) \notag} \\ 
&&{ \qquad = \log l(\mathbf{Z}_{k}|X_{k} = X_{\text{true}, k} , \boldsymbol{\alpha}_{k}, \Delta\mathbf{t} = \Delta\mathbf{t}_{\text{true}}),}
\label{eqn:log_likelihood_with_truth} 
\end{eqnarray} 
{where $ X_{\text{true}, k}$ and $\Delta\mathbf{t}_{\text{true}}$ are the true values of $X_{k}$ and $\Delta\mathbf{t}$.}  

{As a result, the Fishier information of $\mathbf{I}(\alpha_{m,k})$ in $\text{\eqref{eqn:Fisher_Information}}$ is given by}
\begin{eqnarray}
{\mathbf{I}(\alpha_{m,k}) = \mspace{-20mu}\sum_{r\in {\cal E}_m(X_{\text{true},k})}\mspace{-20mu}2\mathbf{s}_m(r,X_{\text{true},k})^{H}\Sigma_m^{-1}\mathbf{s}_m(r,X_{\text{true},k}),} 
\label{eqn:Fisher_information_with_truth}
\end{eqnarray}
{and, the CRB for the $m$th reflection coefficient at the $k$th CPI is found by using the inverse $\mathbf{I}(\alpha_{m,k})$, i.e.,}
\begin{eqnarray}
{\sigma^{2}_{\text{CRB}} \triangleq \mathbf{I}(\alpha_{m,k})^{-1}.} 
\label{eqn:CRLB_local}
\end{eqnarray}
{This quantity is the lower bound of the variance of the complex reflection coefficient, i.e.,} 
\begin{equation}
{\mathrm{Var}(\hat{\alpha}_{m,k}) \geq \sigma^{2}_{\text{CRB}},}
\end{equation}
{where $\mathrm{Var}(\hat{\alpha}_{m,k}) = \mathrm{E}\{|\alpha_{m,k}-\hat{\alpha}_{m,k}|^2\}$ is the variance.}

{Note that $\Sigma_m$ is Hermitian and positive definite. Therefore, the CRB for the real part of the complex reflection is equivalent to that for the imaginary part~$\text{\cite[Chp.15]{Kay1993}}$.}

\ifCLASSOPTIONcaptionsoff
  \newpage
\fi




\begin{thebibliography}{10}
\providecommand{\url}[1]{#1}
\csname url@samestyle\endcsname
\providecommand{\newblock}{\relax}
\providecommand{\bibinfo}[2]{#2}
\providecommand{\BIBentrySTDinterwordspacing}{\spaceskip=0pt\relax}
\providecommand{\BIBentryALTinterwordstretchfactor}{4}
\providecommand{\BIBentryALTinterwordspacing}{\spaceskip=\fontdimen2\font plus
\BIBentryALTinterwordstretchfactor\fontdimen3\font minus
  \fontdimen4\font\relax}
\providecommand{\BIBforeignlanguage}[2]{{%
\expandafter\ifx\csname l@#1\endcsname\relax
\typeout{** WARNING: IEEEtran.bst: No hyphenation pattern has been}%
\typeout{** loaded for the language `#1'. Using the pattern for}%
\typeout{** the default language instead.}%
\else
\language=\csname l@#1\endcsname
\fi
#2}}
\providecommand{\BIBdecl}{\relax}
\BIBdecl

\bibitem{RichardsMelvinScheerEtAl2014}
M.~Richards, W.~Melvin, J.~Scheer, J.~Scheer, and W.~Holm, \emph{Principles of
  Modern Radar: Radar Applications}, ser. Electromagnetics and Radar.\hskip 1em
  plus 0.5em minus 0.4em\relax Institution of Engineering and Technology, 2014.

\bibitem{Haykin2006}
S.~Haykin, ``Cognitive radar: a way of the future,'' \emph{IEEE Signal
  Processing Magazine}, vol.~23, no.~1, pp. 30--40, Jan 2006.

\bibitem{Richards2005b}
M.~Richards, \emph{Fundamentals of Radar Signal Processing}, ser. Professional
  Engineering.\hskip 1em plus 0.5em minus 0.4em\relax Mcgraw-hill, 2005.

\bibitem{Chen2014}
X.~Chen, J.~Guan, N.~Liu, and Y.~He, ``Maneuvering target detection via
  radon-fractional fourier transform-based long-time coherent integration,''
  \emph{Signal Processing, IEEE Transactions on}, vol.~62, no.~4, pp. 939--953,
  Feb 2014.

\bibitem{KongLiCuiEtAl2015}
L.~Kong, X.~Li, G.~Cui, W.~Yi, and Y.~Yang, ``Coherent integration algorithm
  for a maneuvering target with high-order range migration,'' \emph{IEEE
  Transactions on Signal Processing}, vol.~63, no.~17, pp. 4474--4486, Sept
  2015.

\bibitem{Boers2004}
Y.~Boers and J.~Driessen, ``Multitarget particle filter track before detect
  application,'' \emph{Radar, Sonar and Navigation, IEE Proceedings}, vol. 151,
  no.~6, pp. 351--357, Dec 2004.

\bibitem{Grossi2013a}
E.~Grossi, M.~Lops, and L.~Venturino, ``A novel dynamic programming algorithm
  for track-before-detect in radar systems,'' \emph{Signal Processing, IEEE
  Transactions on}, vol.~61, no.~10, pp. 2608--2619, May 2013.

\bibitem{VanTrees1992b}
H.~L. Van~Trees, \emph{Detection, Estimation, and Modulation Theory:
  Radar-Sonar Signal Processing and Gaussian Signals in Noise}.\hskip 1em plus
  0.5em minus 0.4em\relax Melbourne, FL, USA: Krieger Publishing Co., Inc.,
  1992.

\bibitem{Davey2012}
S.~Davey, M.~Rutten, and B.~Cheung, ``Using phase to improve
  track-before-detect,'' \emph{Aerospace and Electronic Systems, IEEE
  Transactions on}, vol.~48, no.~1, pp. 832--849, Jan 2012.

\bibitem{Uney2015}
M.~Uney, B.~Mulgrew, and D.~Clark, ``Maximum likelihood signal parameter
  estimation via track before detect,'' pp. 1--5, Sept 2015.

\bibitem{Rabaste2012}
O.~Rabaste, C.~Riche, and A.~Lepoutre, ``Long-time coherent integration for low
  {SNR} target via particle filter in track-before-detect,'' in
  \emph{Information Fusion (FUSION), 2012 15th International Conference on},
  July 2012, pp. 127--134.

\bibitem{KimUneyMulgrew2016}
K.~Kim, M.~Uney, and B.~Mulgrew, ``Detection of manoeuvring low {SNR} objects
  in receiver arrays,'' in \emph{2016 Sensor Signal Processing for Defence
  (SSPD)}, Sept 2016, pp. 1--5.

\bibitem{KimUeneyMulgrew2017}
------, ``Simultaneous tracking and long time integration for detection in
  collaborative array radars,'' in \emph{2017 IEEE Radar Conference
  (RadarConf)}, May 2017, pp. 0200--0205.

\bibitem{Godrich2010}
H.~Godrich, A.~Haimovich, and R.~Blum, ``Target localization accuracy gain in
  {MIMO} radar-based systems,'' \emph{Information Theory, IEEE Transactions
  on}, vol.~56, no.~6, pp. 2783--2803, June 2010.

\bibitem{Niu2012}
R.~Niu, R.~Blum, P.~Varshney, and A.~Drozd, ``Target localization and tracking
  in noncoherent multiple-input multiple-output radar systems,''
  \emph{Aerospace and Electronic Systems, IEEE Transactions on}, vol.~48,
  no.~2, pp. 1466--1489, April 2012.

\bibitem{Moon1996a}
T.~K. Moon, ``The {E}xpectation-{M}aximization {A}lgorithm,'' \emph{IEEE Signal
  Processing Magazine}, vol.~13, no.~6, pp. 47--60, Nov 1996.

\bibitem{CarlinLouis2010}
B.~Carlin and T.~Louis, \emph{Bayes and Empirical Bayes Methods for Data
  Analysis, Second Edition}, ser. Chapman \& Hall/CRC Texts in Statistical
  Science.\hskip 1em plus 0.5em minus 0.4em\relax Taylor \& Francis, 2010.

\bibitem{Haimovich2008}
A.~Haimovich, R.~Blum, and L.~Cimini, ``{MIMO} radar with widely separated
  antennas,'' \emph{Signal Processing Magazine, IEEE}, vol.~25, no.~1, pp.
  116--129, 2008.

\bibitem{Li2009}
J.~Li and P.~Stoica, \emph{{MIMO} Radar Signal Processing}.\hskip 1em plus
  0.5em minus 0.4em\relax John Wiley \& Sons, Inc., Hoboken, NJ, 2009.

\bibitem{DeMaio2007}
A.~De~Maio and M.~Lops, ``Design principles of {MIMO} radar detectors,''
  \emph{Aerospace and Electronic Systems, IEEE Transactions on}, vol.~43,
  no.~3, pp. 886--898, July 2007.

\bibitem{YangBlum2007}
Y.~Yang and R.~S. Blum, ``{MIMO} radar waveform design based on mutual
  information and minimum mean-square error estimation,'' \emph{IEEE
  Transactions on Aerospace and Electronic Systems}, vol.~43, no.~1, pp.
  330--343, January 2007.

\bibitem{VanTrees2002m}
H.~L. Van~Trees, \emph{Optimum Array Processing}.\hskip 1em plus 0.5em minus
  0.4em\relax New York: Wiley-Interscience, 2002.

\bibitem{Kay1998}
S.~Kay, \emph{Fundamentals of Statistical Signal Processing: Detection theory},
  ser. Prentice Hall Signal Processing Series.\hskip 1em plus 0.5em minus
  0.4em\relax Prentice-Hall PTR, 1998.

\bibitem{B.RisticGordon2004}
S.~A. B.~Ristic and N.~Gordon, \emph{Beyond the Kalman Filter: Particle Filters
  for Tracking Applications}.\hskip 1em plus 0.5em minus 0.4em\relax Artech
  House, 2004.

\bibitem{YaakovBar-Shalom2001}
T.~K. Yaakov Bar-Shalom, X. Rong~Li, \emph{Estimation with Applications To
  Trancking and Navigation: Theory algorithm and Software}.\hskip 1em plus
  0.5em minus 0.4em\relax John Wiley \& Sons, 2001.

\bibitem{Murphy}
K.~Murphy, \emph{Machine Learning: A Probabilistic Perspective}, ser. Adaptive
  computation and machine learning.\hskip 1em plus 0.5em minus 0.4em\relax MIT
  Press.

\bibitem{OrtonFitzgerald2002}
M.~Orton and W.~Fitzgerald, ``A {B}ayesian approach to tracking multiple
  targets using sensor arrays and particle filters,'' \emph{Signal Processing,
  IEEE Transactions on}, vol.~50, no.~2, pp. 216--223, Feb 2002.

\bibitem{Arulampalam2002}
M.~Arulampalam, S.~Maskell, N.~Gordon, and T.~Clapp, ``A tutorial on particle
  filters for online nonlinear/non-gaussian bayesian tracking,'' \emph{Signal
  Processing, IEEE Transactions on}, vol.~50, no.~2, pp. 174--188, Feb 2002.

\bibitem{Casella2005}
G.~Casella and C.~P. Robert, \emph{Monte Carlo Statistical Methods},
  2nd~ed.\hskip 1em plus 0.5em minus 0.4em\relax Springer, 2005.

\bibitem{Bazaraa1993}
H.~D.~S. M.~S.~Bazaraa and C.~Shetty, \emph{Nonlinear Programming},
  2nd~ed.\hskip 1em plus 0.5em minus 0.4em\relax John Wiley \& Sons, Inc.,
  1993.

\bibitem{Kay1993}
S.~M. Kay, \emph{Fundamentals of Statistical Signal Processing: Estimation
  Theory}.\hskip 1em plus 0.5em minus 0.4em\relax Upper Saddle River, NJ, USA:
  Prentice-Hall, Inc., 1993.

\end{thebibliography}
\bibliographystyle{IEEEtran}
\end{document}